\RequirePackage{fix-cm}

\documentclass[twocolumn]{svjour3}          
\smartqed  
\usepackage{graphicx}
\usepackage{enumitem}
\usepackage{natbib}
\usepackage{amsmath,amssymb}
\usepackage{cite}
\usepackage[ruled]{algorithm2e} 
\usepackage{qtree}
\usepackage{xcolor}
\usepackage{algpseudocode}
\usepackage{placeins}

\usepackage{appendix}

\journalname{Social Network Analysis and Mining}
\begin{document}

\title{A Query-Driven System for Discovering Interesting Subgraphs in Social Media
}

\titlerunning{Subgraph Discovery}        

\author{Subhasis Dasgupta         \and
        Amarnath Gupta 
}
\institute{A. Gupta \at
              San Diego Supercomputer Center, University of California San Diego \\
              \email{a1gupta@ucsd.edu}           
           \and
           Subhasis Dasgupta \at
            San Diego Supercomputer Center, University of California San Diego \\
            \email{sudasgupta@ucsd.edu}
}
\date{Received: date / Accepted: date}
\twocolumn[
  \begin{@twocolumnfalse}
    \maketitle
    \begin{abstract}
Social media data are often modeled as heterogeneous graphs with multiple types of nodes and edges. We present a discovery algorithm that first chooses a ``background'' graph based on a user's analytical interest and then automatically discovers subgraphs that are structurally and content-wise distinctly different from the background graph. The technique combines the notion of a \texttt{group-by} operation on a graph and the notion of subjective interestingness, resulting in an automated discovery of interesting subgraphs. Our experiments on a socio-political database show the effectiveness of our technique.
    \end{abstract}
  \end{@twocolumnfalse}
]
\maketitle
\section{Introduction}
\label{sec:intro}
An information system designed for analysis of social media must consider a common set properties that characterize all social media data.
\begin{itemize}[leftmargin=*]
    \item Information elements in social media are essentially \textit{heterogeneous} in nature -- users, posts, images, external URL references, although related, all bear different kinds information. 
    \item Most social information is \textit{temporal} -- a timestamp is associated with user events like the creation or response on a post, as well as system events like user account creation, deactivation and deletion. The system should therefore allow both temporal as well as time-agnostic analyses.
    \item Information in social media evolves fast. In one study \citep{zhu2013modeling}, it was shown that the number of users in a social media is a power function of time. More recently, \citep{antonakaki2018utilizing} showed that Twitter's growth is supralinear and follows Lescovec's model of graph evolution \citep{leskovec2007graph}. Therefore, an analyst may first have to perform \textit{exploration tasks} on the data before figuring out their analysis plan. 
    \item Social media has a significant textual content, sometimes with specific entity markers (e.g., mentions) and topic markers (e.g., hashtags). Therefore any information element derived from text (e.g., named entities, topics, sentiment scores) may also be used for analysis. To be practically useful, the system must accommodate semantic synonyms -- \#KamalaHarris, @KamalaHarris and ``Kamala Harris'' refer to the same entity.
    \item Relationships between information items in social media data must capture both canonical relationships like (\texttt{tweet-15 mentions user-392}) but a wide-variety of computed relationships over base entities (users, posts, $\ldots$) and text-derived information (e.g., named entities). 
\end{itemize}
It is also imperative that such an information must support three styles of analysis tasks 
\begin{enumerate}
    \item \textbf{Search}, where the user specifies content predicate without specifying the structure of the data. For example, seeking the number of tweets related to Kamala Harris should count tweets where she is the author, as well as tweets where any synonym of ``Kamala Haris'' is in the tweet text.
    \item \textbf{Query}, where the user specifies query conditions based on the structure of the data. For example, tweets with \texttt{create\_date} between 9 and 9:30 am on January 6th, 2021, with \texttt{text} containing the string ``Pence'' and that were \texttt{favorited} at least 100 times during the same time period.
    \item \textbf{Discovery,} where the user may or may not know the exact predicates on the data items to be retrieved, but can specify analytical operations (together with some post-filters) whose results will provide insights into the data. For example, we call a query like \texttt{Perform \textit{community detection} on all tweets on January 6, 2021 and return the users from the largest community} a discovery query.
\end{enumerate}
In general, a real-life analytics workload will freely combine these modalities as part of a user's information exploration process. 

In this paper, we present a general-purpose graph-based  model for social media data and a subgraph discovery algorithm atop this data model. Physically, the data model is implemented on AWESOME \citep{gupta:awesome:2016}, an analytical platform designed to enable large-scale social media analytics over continuously acquired data from social media APIs. The platform, developed as a polystore system natively supports relational, graph and document data, and hence enables a user to perform complex analysis that include arbitrary combinations of search, query and discovery operations. We use the term \textbf{\textit{query-driven discovery}} to reflect that the scenario where the user does not want to run the discovery algorithm on a large and continuously collected body of data; rather, the user knows a starting point that can be specified as an expressive query (illustrated later), and puts bounds on the discovery process so that it terminates within an acceptable time limit.

\medskip 

\noindent \textbf{Contributions.} This paper makes the following contributions. (a) It offers a new formulation for the subgraph interestingness problem for social media; (b) based on this formulation, it presents a discovery algorithm for social media; (c) it demonstrates the efficacy of the algorithm on multiple data sets.

\medskip

\noindent \textbf{Organization of the paper.} The rest of the paper is organized as follows. Section \ref{sec:related} describes the related research on interesting subgraph finding in as investigated by researchers in Knowledge Discovery, Information Management, as well Social Network Mining. Section \ref{sec:prelim} presents the abstract data model over which the Discovery Algorithm operations and the basic definitions to establish the domain of discourse for the discovery process. Section \ref{sec:generate} presents our method of generating candidate subgraphs that will be tested for interestingness. Section \ref{sec:discovery} first presents our interestingness metrics and then the testing process based on these metrics. Section \ref{sec:experiments} describes the experimental validation of our approach on multiple data sets. Section \ref{sec:conclusion} presents concluding discussions.

\section{Related Work}
\label{sec:related}
The problem of finding ``interesting'' information in a data set is not new. \citep{what-makes-patterns-interesting:1996} described that  an``interestingness measure'' can be ``objective'' or ``subjective''. A measure is ``objective'' when it is computed solely based on the properties of the data. In contrast, a ``subjective'' measure must take into account the user's perspective. They propose that (a) a pattern is interesting if it is "surprising" to the user (\textit{unexpectedness}) and (b) a pattern is interesting if the user can act on it
to his advantage (\textit{actionability}). Of these criteria, actionability is hard to determine algorithmically; unexpectedness, on the other hand, can be viewed as the departure from the user's beliefs. For example, a user may believe that the 24-hour occurrence pattern of all hashtags are nearly identical. In this case, a discovery would be to find a set of hashtags and sample dates for which this belief is violated. 
Following \citep{geng2006interestingness}, there are three possibilities regarding how a system is informed of a user's knowledge and beliefs: (a) the user provides a formal specification of his or her knowledge, and after obtaining the mining results, the system chooses which unexpected patterns to present to the user \citep{BingLiu:1999}; (b) according to the user’s interactive feedback, the system removes uninteresting patterns \citep{not-interesting:1999}; and (c) the system applies the user’s specifications as constraints during the mining process to narrow down the search space and provide fewer results. Our work roughly corresponds to the third strategy.
\begin{figure*}[t]
    \centering
    \begin{minipage}{.5\textwidth}
    \centering
      \includegraphics[width=6.5cm, height=6.5cm]{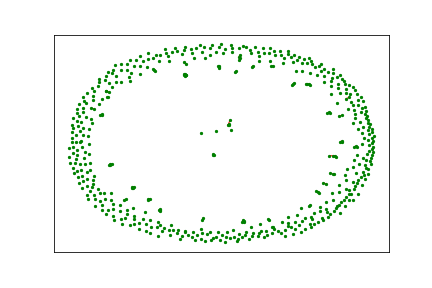}

    \end{minipage}%
    \begin{minipage}{.5\textwidth}
     \centering
     \includegraphics[width=6.5cm, height=6.5cm]{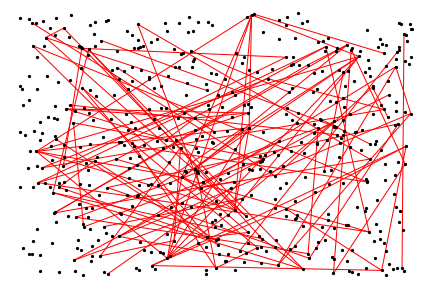}
       
    \end{minipage}
    \begin{minipage}{.4\textwidth}
    \centering
        \includegraphics[width=7.5cm, height=6.5cm]{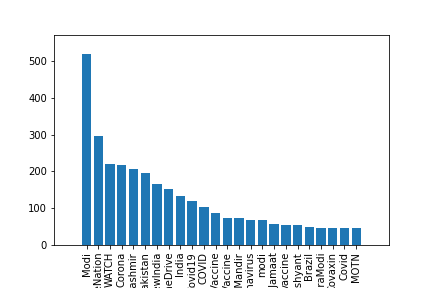}
    \end{minipage}
    \begin{minipage}{.4\textwidth}
    \centering
        \includegraphics[width=7.5cm, height=6.5cm]{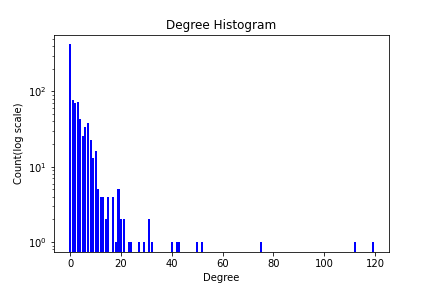}
    \end{minipage}
    \caption{An example of an interesting subgraph. Tweets on Indian politics form a sparse center in a dense periphery (separately shown in the top right). The bottom two figures show the hashtag distribution of the center and the degree distribution of the center (log scale) respectively.}
    \label{fig:sparse}
\end{figure*}
Early research on finding interesting subgraphs focused primarily on finding interesting substructures. This body of research primarily found interestingness in two directions: (a) finding frequently occurring subgraphs in a collection of graphs (e.g., chemical structures) \citep{kuramochi2001frequent, yan2003closegraph, thoma2010discriminative} and (b) finding regions of a large graph that have high edge density \citep{lee2010survey, sariyuce2015finding, epasto2015efficient, wen2017efficient} compared to other regions in the graph. Note that while a dense region in the graph can definitely interesting, shows, the inverse situation where a sparsely connected region is surrounded by an otherwise dense periphery can be equally interesting for an application. 

We illustrate the situation in Figure \ref{fig:sparse}. The primary data set is a collection of tweets on COVID-19 vaccination, but this specific graph shows a sparse core on Indian politics that is loosely connected to nodes of an otherwise dense periphery on the primary topic. Standard network features like hashtag histogram and the node degree histograms do not reveal this substructure requiring us to explore new methods of discovery. 

A serious limitation of the above class of work is that the interestingness criteria does not take into account \textit{node content} (resp. edge content) which may be present in a property graph data model \citep{angles2018property} where nodes and edges have attributes. \citep{bendimerad2019mining} presents several discovery algorithms for graphs with vertex attributes and edge attributes. They perform both structure-based graph clustering and subspace clustering of attributes to identify interesting (in their domain ``anomalous'') subgraphs.

On the ``subjective'' side of the interestingness problem, one approach considers interesting subgraphs as a subgraph matching problem \citep{shan2019dynamic}. Their general idea is to compute all matching subgraphs that satisfy a user the query and then ranking the results based on the rarity and the likelihood of the associations among entities in the subgraphs. In contrast \citep{adriaens2019subjectively} uses the notion of ``subjective interestingness'' which roughly corresponds to finding subgraphs whose connectivity properties (e.g., the average degree of a vertices) are distinctly different from an ``expected'' \textit{background} graph. This approach uses a constrained optimization problem that maximizes an objective function over the \textit{information content} ($IC$) and the \textit{description length} ($DL$) of the desired subgraph pattern. 

Our work is conceptually most inspired by \citep{bendimerad2019subj} that explores the subjective interestingness problem for attributed graphs. Their main contribution centers around CSEA (Cohesive Subgraph with Exceptional Attributes) patterns that inform the user that a given set of attributes has exceptional values throughout a set of vertices in the graph. The subjective interestingness is given by $$S(U,S) = \frac{IC(U,S)}{DL(U)}$$ where $U$ is a subset of nodes and $S$ is a set of restrictions on the value domains of the attributes. The system models the prior beliefs of the user as the Maximum Entropy distribution subject to any stated prior, beliefs the user may hold about the data (e.g., the distribution of an attribute value). The information content $IC(U,S)$ of a CSEA pattern $(U,S)$ is  formalized as negative of the logarithm of the probability that the pattern is present under the background distribution.  The length of a description
of $U$ is the intersection of all neighborhoods in a subset $X \subseteq N (U)$,
along with the set of ``exceptions'', vertices are in the intersection but not part of
$U$. However, we have a completely different, more database-centric formulation of the background and the user's beliefs.

\section{The Problem Setup}
\label{sec:prelim}

\subsection{Data Model}
\label{sec:dataModel}
Our abstract model social media data takes the form of a \textit{heterogeneous information network} (an information network with multiple types of nodes and edges), which we view as a temporal property graph $G$. Let $N$ be the node set and $E$ be the edge set of $G$. $N$ can be viewed as a disjoint union of different subsets (called \textit{node types}) --  users $U$, posts $P$, topic markers (e.g., hastags) $H$, term vocabulary $V$ (the set of all terms appearing in a corpus), references (e.g., URLs) $R(\tau)$, where $\tau$ represents the type of resource (e.g., image, video, web site $\ldots$). Each type of node have a different set of properties (attributes) $\bar{A}(.)$. We denote the attributes of $U$ as $\bar{A}(U) = a_1(U), a_2(U) \ldots$ such that $a_i(U)$ is the $i$-the attribute of $U$. An attribute of a node type may be temporal -- a post $p \in P$ may have temporal attribute called \texttt{creationDate}. Edges in this network can be directional and have a single \textit{edge type}. The following is a set of \textit{base (but not exhaustive) edge types}:
\begin{itemize}[leftmargin=1em, label=\scriptsize{$-$}]
    \item \texttt{writes}: $U \mapsto P$
    \item \texttt{uses}: $P \mapsto H$
    \item \texttt{mentions}: $P \mapsto U$ maps a post $p$ to a user $u$ if $u$ mentioned in $p$
    \item \texttt{repostOf}: $P \mapsto P$ maps a post $p_2$ to a post $p_1$ if $p_2$ is repost of $p_1$. This implies that $ts(p_2) < ts(p_1)$ where $ts$ is the timestamp attribute
    \item \texttt{replyTo/comment}: $P \mapsto P$ maps a post $p_2$ to a post $p_1$ if $p_2$ is a reply to $p_1$. This implies that $ts(p_2) < ts(p_1)$ where $ts$ is the timestamp attribute
    \item \texttt{contains}: $P \mapsto V \times \mathcal{N}$ where $\mathcal{N}$ is the set of natural numbers and represents the count of a token $v \in V$ in a post $p \in P$
\end{itemize}
 We realistically assume that the inverse of these mappings can be computed, i.e., if $v_1, v_2 \in V$ are terms, we can perform a \texttt{contains}$^{-1}(v_1 \wedge \neg v_2)$ operation to yield all posts that used $v_1$ but not $v_2$.
 
The AWESOME information system allows users construct \textit{derived or computed edge types} depending on the specific discovery problem they want to solve. For example, they can construct a standard hashtag co-occurrence graph using a non-recursive aggregation rule in Datalog \citep{consens1993low}:
\begin{equation*}
    \begin{aligned}
        HC(h_1, h_2, count(p)) &\longleftarrow \\
        &p \in P, h_1 \in H, h_2 \in H \\ 
        &uses(p, h_1), uses(p, h_2) 
    \end{aligned}
\end{equation*}
We interpret $HC(h_1, h_2, count(p))$ as an edge between nodes $h_1$ and $h_2$ and $count(p)$ as an attribute of the edge. In our model, a computed edge has the form: $E_T(N_b, N_e, \bar{B})$ where $E_T$ is the edge type, $N_b, N_e$ are the tail and head nodes of the edge, and $\bar{B}$ designates a flat schema of edge properties. The number of such computed edges can be arbitrarily large and complex for different subgraph discovery problems. A more complex computed edge may look like: $UMUHD(u_1,u_2,mCount,d, h)$, where an edge from $u_1$ to $u_2$ is constructed if user $u_1$ creates a post that contains hashtag $h$, and mentions user $u_2$ a total number of $mCount$ times on day $d$. Note that in this case, hashtag $h$ is a property of edge type $UMUHD$ and is not a topic marker node of graph $G$. 

\subsection{Specifying User Knowledge}
\label{sec:hquery}
Since the discovery process is query-driven, the system has no \textit{a priori} information about the user's interests, prior knowledge and expectations, if any, for a specific discovery task. Hence, given a data corpus, the user needs to provide this information through a set of parameterized queries. We call these queries ``heterogeneous'' because they can be placed on conditions on any arbitrary node (resp. edge) properties, network structure and text properties of the graph. 

\medskip

\noindent \textbf{User Interest Specification.} The discovery process starts with the specification of the user's universe of discourse identified with a query $Q_0$. We provide some illustrative examples of user interest with queries of increasing complexity.

\noindent \textit{Example 1.}  ``All tweets related to COVID-19 between 06/01/2020 and 07/31/2020 that refer to Hydroxychloroquine''. In this specification, the condition ``related to COVID-19'' amounts to finding tweets containing any $k$ terms from a user-provided list, and the condition on ``Hydroxychloroquine'' is expressed as a fuzzy search.
\begin{figure}
\includegraphics[width=8.5cm, height=6.5cm]{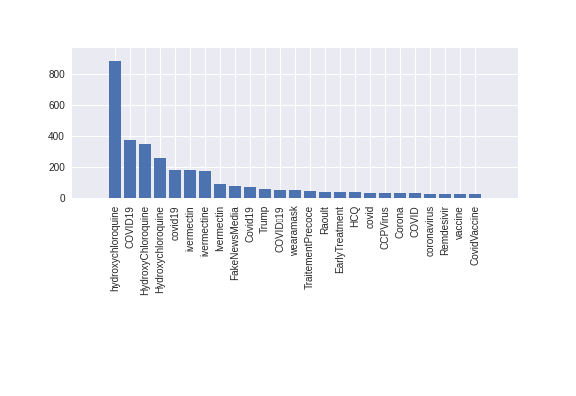}
\vspace{-2cm}
 \caption{Hydroxychloroquine and COVID related Hashtags}
 \label{fig:hydroxi}
\end{figure}
Figure \ref{fig:hydroxi} shows the top hashtags related to this search -- note that the hashtag co-occurrence graph around COVID-19 and hyroxychloroquine includes ``FakeNewsMedia''.

\noindent \textit{Example 2.} ``All tweets from users who mention \texttt{Trump} in their user profile and \texttt{Fauci} in at least $n$ of their tweets''. Notice that this query is about users with a certain behavioral pattern -- it captures \textit{all} tweets from users who have a combination of specific profile features and tweet content.  

\noindent \textit{Example 3.} ``All tweets from users $U_1$ whose tweets appear in hashtag-cooccurrence in the 3 neighborhood around \texttt{\#ados} is used together with all tweets of users $U_2$ who belong to the user-mention-user networks of these users (i.e., $U_1$) '', where \texttt{\#ados} refers to ``American Descendant of Slaves'', which represents an African American cause.

\begin{figure*}[t]
    \centering
    \begin{minipage}{.5\textwidth}
    \centering
      \includegraphics[width=7.5cm, height=6.5cm]{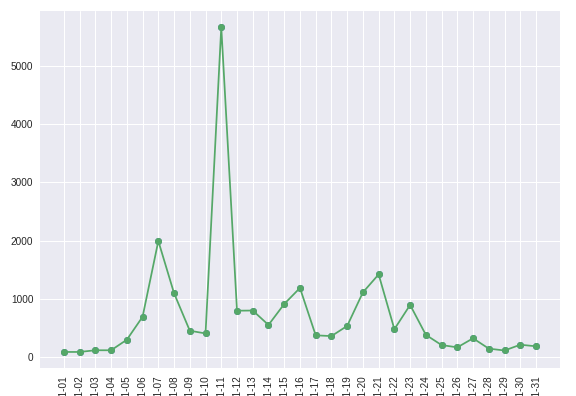}

    \end{minipage}%
    \begin{minipage}{.5\textwidth}
     \centering
     \includegraphics[width=7.5cm, height=6.5cm]{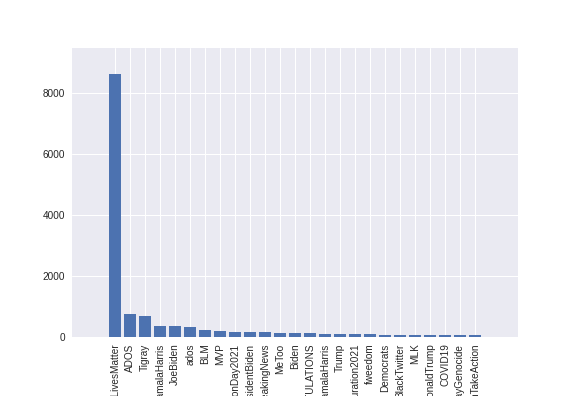}
       
    \end{minipage}

    \caption{Two views of the \texttt{\#ados} cluster of hashtags }
    \label{fig:ados}
\end{figure*}

\noindent The end result of $Q_0$ is a collection of posts that we call the \textit{initial post set} $P_0$. Using the posts in $P_0$, the system creates a background graph as follows.

\medskip

\noindent \textbf{Initial Background Graph.} The initial background graph $G_0$ is the graph derived from $P_0$ over which the discovery process runs. However, to define the initial graph, we first develop the notion of a \textit{conversation}.

\begin{definition}[\textbf{Semantic Neighborhood}.] $N(p)$, the semantic neighborhood of a post $p$ is the graph connecting $p$ to instances of $U \cup H \cup P \cup V$ that directly relates to $p$.
\end{definition}

\begin{definition}[\textbf{Conversation Context}.] The conversation context $C(p)$ of post $p$ is a subgraph satisfying the following conditions:
\begin{enumerate}[leftmargin=*]
    \item $P_1$: The set of posts reachable to/from $p$ along the relationships \texttt{repostOf, replyTo} belong to $C(p)$. 
    \item $P_2$: The union of posts in the semantic neighborhood of $P_1$ belong to $C(p)$.
    \item $E$: The induced subgraph of $P_1 \cup P_2$ belong to $C(p)$
    \item Nothing else belongs to $C(p)$.
\end{enumerate}
\end{definition}
\noindent Clearly, we can assert that $C(p)$ is a connected graph and that $N(p)~ \sqsubset_g~ C(p)$ where $\sqsubset_g$ denotes a subgraph relationship.

\begin{definition}[\textbf{Initial Background Graph}.] The initial background graph $G_0$ is a merger of all conversation contexts $C(p_i), p_i \in P_0$, together with all computed edges induces the nodes of $\cup_i C(p_i)$
\end{definition}
The initial background graph itself can be a gateway to finding interesting properties of the graph. To illustrate this based on the graph obtained from our Example 2. Figure \ref{fig:ados} presents two views of the \texttt{\#ados} cluster of hashtags from January 2021. The left chart shows the time vs. count of the hashtags while the right chart shows the dominant hashtags of the same period in this cluster. The strong peak in the timeline, was due to an intense discussion, revealed by topic modeling, on the creation of an office on African American issues. The occurrence of this peak is interesting because most of the social media conversation in this time period was focused on the Capitol attack on January 6.

Given the $G_0$ graph,we discover subgraphs $S_i \subset G_o$ whose content and structure are distinctly different that of $G_0$. However, unlike previous approaches, we apply a generate-and-test paradigm for discovery. The generate-step (Section \ref{sec:generate}) uses a graph cube like \citep{zhao2011graph} technique to generate candidate subgraphs that might be interesting and the test-step (Section \ref{sec:testing}) computes if (a) the candidate is sufficiently distinct from  the $G'$, and (b) the collection of candidates are sufficiently distinct from each other.

\noindent \textbf{Subgraph Interestingness.} For a subgraph $S_i$ to be considered as a candidate, it must satisfy the following conditions.

\noindent \textbf{C1.} $S_i$ must be connected and should satisfy a size threshold $\theta_n$, the minimal number of nodes.

\noindent \textbf{C2.} Let $A_{ij}$ (resp. $B_{ik}$) be the set of \textit{local} properties of node $j$ (resp. edge $k$) of subgraph $S_i$. A property is called ``local'' if it is not a network property like vertex degree. All nodes (resp. edges) of $S_i$ must satisfy some user-specified predicate $\phi_N$ (resp. $\phi_E$) specified over $A_{ij}$ (resp. $B_{ik}$). For example, a node predicate might require that all ``post'' nodes in the subgraph must have a re-post count of at least 300, while an edge predicate may require that all hashtag co-occurrence relationships must have a weight of at least 10. A user defined constraint on the candidate subgraph improves the interpretability of the result. Typical subjective interestingness techniques \citep{van2016subjective, adriaens2019subjectively} use only structural features of the network and do not consider attribute-based constraints, which limits their pragmatic utility.

\noindent \textbf{C3.} For each text-valued attribute $a$ of $A_{ij}$, let $C(a)$ be the collection of the values of $a$ over all nodes of $S_i$, and $\mathcal{D}(C(a))$ is a textual diversity metric computed over $C(a)$. For $S_i$ to be interesting, it must have at least one attribute $a$ such that $\mathcal{D}(C(a))$ does not have the usual power-law distribution expected in social networks. Zheng et al \citep{zheng2019social} used \textit{vocabulary diversity} and \textit{topic diversity} as textual diversity measures.

\section{Candidate Subgraph Generation}
\label{sec:generate}
Section \ref{sec:hquery} describes the creation of the initial background graph $G_0$ that serves as the domain of discourse for discovery. Depending on the number of initial posts $P_0$ resulting form the initial query, the size of $G_0$ might be too large -- in this case the user can specify followup queries on $G_0$ to narrow down the scope of discovery. We call this narrowed-down graph of interest as $G'$ -- if no followup queries were used, $G' = G_0$. The next step is to generate some candidate subgraphs that will be tested for interestingness. 

\noindent \textbf{Node Grouping.} A node group is a subset of \textit{nodes($G'$)} where all nodes in a group have some similar property. We generalize the \textit{groupby} operation, commonly used in relational database systems, to heterogeneous information networks. To describe the generalization, let us assume $R(A, B, C, D, \ldots)$ is a relation (table) with attributes $A, B, C, D, \ldots$ A \textit{groupby} operation takes as input (a) a subset of \textit{grouping attributes} (e.g. $A, B$), (b) a \textit{grouped attribute} (e.g., $C$) and (c) an \textit{aggregation function} (e.g., \textit{count}). The operation first computes each distinct cross-product value of the grouping attributes (in our example, $A \times B$) and creates a list of all values of the grouped attribute corresponding to each distinct value of the grouping attributes, and then applies the aggregation function to the list. Thus, the result of the \textit{groupby} operation is a single aggregated value for each distinct cross-product value of grouping attributes. 

To apply this operation to a social network graph, we recognize that there are two distinct ways of defining the ``grouping-object''. \\
(1) Node properties can be directly used just like in the relational case. For example, for tweets a grouping condition might be \texttt{getDate} \texttt{(Tweet.created\_at)} $\wedge$ \texttt{bin(Tweet.favoriteCount, 100)}, where the \texttt{getDate} function extracts the date of a tweet and the \texttt{bin} function creates buckets of size 100 from the favorite count of each tweet. \\
(2) The grouping-object is a subgraph pattern. For example, the subgraph pattern\\
\texttt{(:tweet\{date\})-[:uses]->(:hashtag\{text\})} \hfill{(P1)}\\
states that all ''tweet'' nodes having the same posting date, together with every distinct hashtag text will be placed in a separate group. Notice that while (1) produces disjoint tweets, (2) produces a ``soft'' partitioning on the tweets and hashtags due to the many-to-many relationship between tweets and hashtags. \\
In either case, the result is a set of node groups, designated here as $N_i$. For example, the grouping pattern P1 expressed in a Cypher-like syntax \citep{francis2018cypher} (implemented in the Neo4J graph data management system) states that all tweets having the same posting date, together with every distinct hashtag text will be placed in a separate group.
Notice that this process produces a ``fuzzy'' partitioning on the tweets and hashtags due to the many-to-many relationship between tweets and hashtags. Hence, the same tweet node can belong to two different groups because it has multiple hashtags. Similarly, a hashtag node can belong to multiple groups because tweets from different dates may have used the same hashtag. While the grouping condition specification language can express more complex grouping conditions, in this paper, we will use simpler cases to highlight the efficacy of the discovery algorithm. We denote the node set in each group as $N_i$.

\noindent \textbf{Graph Construction.} To complete the \textit{groupby} operation, we also need to specify the aggregation function in addition to the grouping-object and the grouped-object. This function takes the form of a graph construction operation that constructs a subgraph $S_i$ by expanding on the node set $N_i$. Different expansion rules can be specified, leading to the formation of different graphs. Here we list three rules that we have found fairly useful in practice.

\noindent \textbf{G1.} Identify all the \texttt{tweet} nodes in $N_i$. Construct a \textit{relaxed induced subgraph} of the \texttt{tweet}-labeled nodes in $N_i$. The subgraph is induced because it only uses tweets contained within $N_i$, and it is \textit{relaxed} because contains all nodes \textit{directly associated} with these tweet nodes, such as author,  hashtags, URLs, and mentioned-users. 

\noindent \textbf{G2.} Construct a \textit{mention network} from within the tweet nodes in $N_i$ -- the mention network initially connects all \texttt{tweet} and \texttt{user}-labeled nodes. Extend the network by including all nodes \textit{directly associated} with these tweet nodes.

\noindent \textbf{G3.} A third construction relaxes the grouping constraint. We first compute either \textbf{G1} or \textbf{G2}, and then extend the graph by including the first order neighborhood of mentioned users or hashtags. While this clearly breaks the initial group boundaries, a network thus constructed includes tweets of similar themes (through hashtags) or audience (through mentions).

\noindent \textbf{Automated Group Generation.} In a practical setting, as shown in Section \ref{sec:experiments}, the parameters for node grouping operation can be specified by a user, or it can be generated automatically. Automatic generation of grouping-objects is based on the considerations described below. To keep the autogeneration manageable, we will only consider single and two objects for attribute grouping and only a single edge for subgraph patterns.
\begin{itemize}[leftmargin=*]
    \item Since temporal shifts in social media themes and structure are almost always of interest, the posting timestamp is always a grouping variable. For our purposes, we set the granularity to a day by default, although a user can set it.
    \item The frequency of most nontemporal attributes (like hashtags) have a mixture distribution of double-pareto lognormal distribution and power law \citep{gupta:osn:2020}, we will adopt the following strategy.
    \begin{itemize}[label={$\circ$}]
        \item  Let $f(A)$ be distribution of attribute $A$, and $\kappa(f(A))$ be the curvature of $f(A)$. If $A$ is a discrete variable, we find $a*$, the maximum curvature (elbow) point of $f(A)$ numerically \citep{antunes2018knee}.
        \item  We compute $A'$, the values of attribute $A$ to the left of $a*$ for all attributes and choose the attribute where the cardinality of $A'$ is maximum. In other words, we choose attributes which have the highest number of pre-elbow values. 
    \end{itemize}
    \item We adopt a similar strategy for subgraph patterns. If $T_1(a_i)\stackrel{L}{\longrightarrow}T_2(b_j)$ is an edge where $T_1, T_2$ are node labels, $a_i, b_j$ are node properties and $L$ is an edge label, then $a_i$ and $b_j$ will be selected based on the conditions above. Since the number of edge labels is fairly small in our social media data, we will evaluate the estimated cardinality of the edge for all such triples and select one with the lowest cardinality. 
\end{itemize}

\section{The Discovery Process}
\label{sec:discovery}

\subsection{Measures of for Relative Interestingness}
\label{sec:interestingness}
We compute the interestingness of a subgraph $S$ in reference to a background graph $G_b$ (e.g., $G'$), and consists of a structural as well as a content component. We first discuss the structural component. To compare a subgraph $S_i$ with the background graph, we first compute a set of network properties $P_j$ (see below) for nodes (or edges) and then compute the frequency distribution $f(P_j(S_i))$ of these properties over all nodes (resp. edges) of (a) subgraphs $S_i$, and (b) the reference graph (e.g., $G'$). A distance between $f(P_j(S_i))$ and $f(P_j(G_b))$ is computed using Jensen–Shannon divergence (JSD). In the following, we use $\Delta(f_1,f_2)$ to refer to the JS-divergence of distributions $f_1$ and $f_2$. 
\medskip
\noindent \textbf{Eigenvector Centrality Disparity:} The testing process starts by identifying the distributions of nodes with high node centrality between the networks. While there is no shortage of centrality measures in the literature, we choose eigenvector centrality \citep{das2018study} defined below, to represent the dominant nodes. Let $A = (a_{i,j})$ be the adjacency matrix of a graph. The eigenvector centrality $x_{i}$ of node $i$ is given by: $$x_i = \frac{1}{\lambda} \sum_k a_{k,i} \, x_k$$ where $\lambda \neq 0$ is a constant. 
    The rationale for this choice follows from earlier studies in \citep{Bonacich2007-mx,Ruhnau2000-jy,Yan2014-dn}, who establish that since the eigenvector centrality can be seen as a weighted sum of direct and indirect connections, 
    it represents the true structure of the network more faithfully than other centrality measures. 
    Further, \citep{Ruhnau2000-jy} proved that the eigenvector-centrality under the Euclidean norm can be transformed into node-centrality, a property not exhibited by other common measures.
    Let the distributions of eigenvector centrality of subgraphs $A$ and $B$ be $\beta_a$ and $\beta_b$ respectively, and that of the background graph be $\beta_t$, then 
    $|\Delta_e(\beta_t, \beta_a)| > \theta $ indicates that $A$ is sufficiently structurally distinct from $G_b$ 
    $|\Delta_e(\beta_t, \beta_a)| > |\Delta_e(\beta_t, \beta_b)|$ indicates that $A$ contains significantly more or significantly less influential nodes than $B$. 

\medskip
\noindent \textbf{Topical Navigability Disparity:} Navigability measures ease of flow. If subgraph $S$ is more navigable than subgraph $S'$, then there will be more traffic through $S$ compared to $S'$. However, the likelihood of seeing a higher flow through a subgraph depends not just on the structure of the network, but on extrinsic covariates like time and topic. So, a subgraph is interesting in terms of navigability if for some values of a covariate, its navigability measure is different from that of a background subgraph. 

Inspired by its application in biology \citep{seguin2018navigation}, traffic analysis \citep{scellato2010traffic}, and network attack analysis \citep{lekha2020central}, we use \textit{edge betweenness centrality} \citep{das2018study} as the generic (non-topic) measure of navigability.   Let $\alpha_{ij}$ be the number of shortest paths from node i to j and $\alpha_{ij}(k)$ is the number of paths passes through the edge $k$. Then the edge-betweenness centrality is $$C_{eb}(k)= \sum_{(i,j)\in V} \frac{\alpha_{ij}(k)}{\alpha_{ij}}$$
By this definition, the edge betweenness centrality is the portion of all-pairs shortest paths that pass through an edge. Since edge betweenness centrality of edge $e$ measures the proportion of paths that passes through $e$, a subgraph $S$ with a higher proportion of high-valued edge betweenness centrality implies that $S$ may be more \textit{navigable} than $G_b$ or another subgraph $S'$ of the graph, i.e., information propagation is higher through this subgraph compared to the whole background network, for that matter, any other subgraph of network having a lower proportion of nodes with high edge betweenness centrality. Let the distribution of the edge betweenness centrality of two subgraphs $A$ and $B$ are $c_1$ and $c_2$ respectively, and that of the reference graph is $c_0$. Then, $|\Delta_b(c_0, c_1)| < |\Delta_b(c_0, c_2)|$  means the second subgraph is more navigable than the first.

To associate navigability with topics, we detect topic clusters over the background graph and the subgraph being inspected. The exact method for topic cluster finding is independent of the use of topical navigability. In our setting, we have used topic  topic modeling and dense region detection in hashtag cooccurrence networks. For each topic cluster, we identify posts (within the subgraph) that belong to the cluster. If the number of posts is greater than a threshold, we compute the navigability disparity.

\medskip

\begin{figure*}[t]
    \centering
    \begin{minipage}{.5\textwidth}
    \centering
      \includegraphics[width=7.5cm, height=5.5cm]{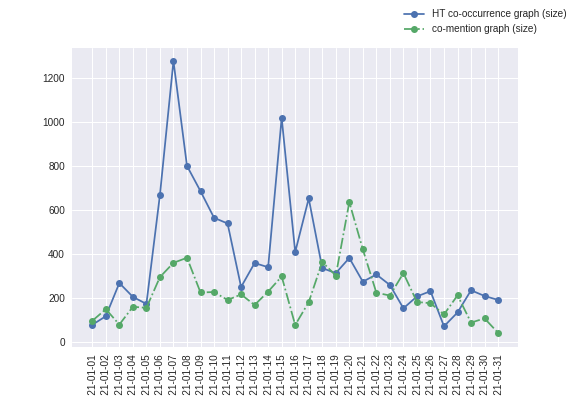}

        \caption{\#ADOS and Black Lives Matter }
    \end{minipage}%
    \begin{minipage}{.5\textwidth}
     \centering
     \includegraphics[width=7.5cm, height=5.5cm]{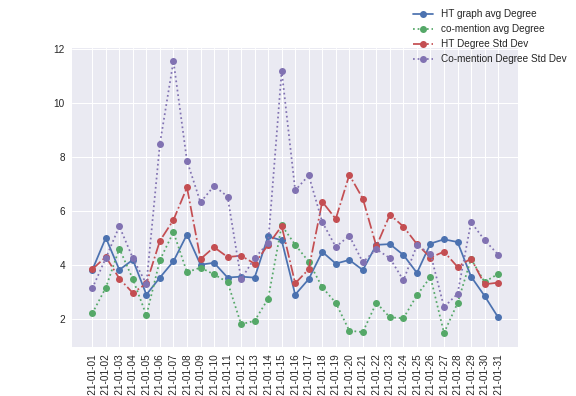}
       
       \caption{AVG Degree Distributions and Std Dev }
    \end{minipage}

     \centering
    \begin{minipage}{.5\textwidth}
    \centering
      \includegraphics[width=7.5cm, height=5.5cm]{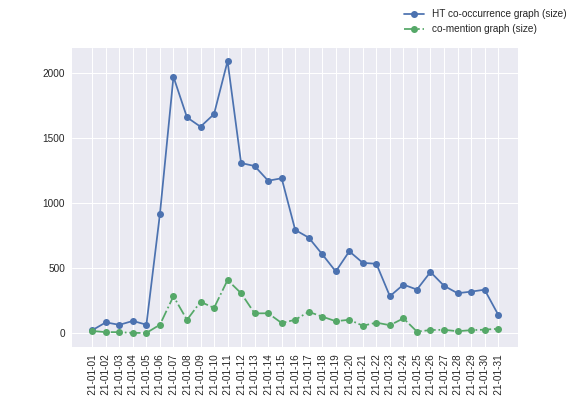}
        \caption{Insurrection and Capitol Attack}
    \end{minipage}%
    \begin{minipage}{.5\textwidth}
     \centering
     \includegraphics[width=7.5cm, height=5.5cm]{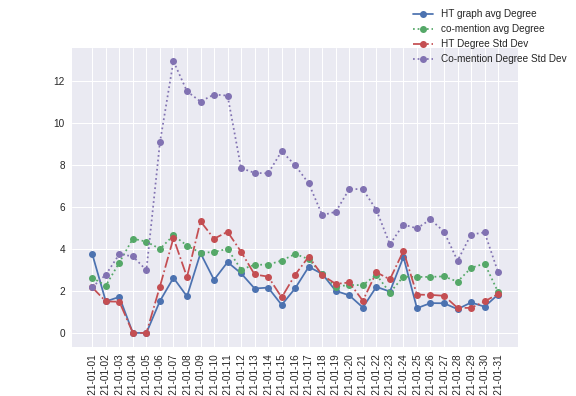}
       
      \caption{AVG Degree Distributions and Std Dev}
    \end{minipage}
        \begin{minipage}{.5\textwidth}
    \centering
      \includegraphics[width=7.5cm, height=5cm]{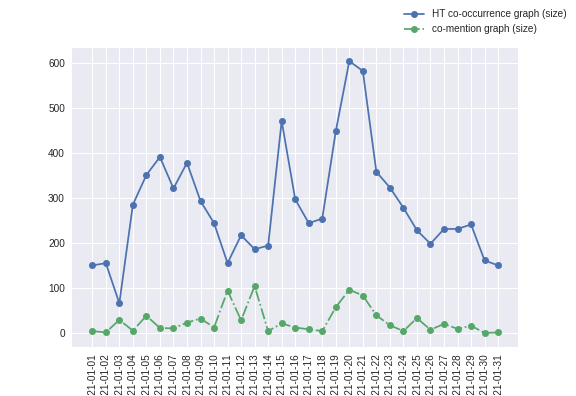}
          \caption{Socioeconomic Issues During COVID-19 }
    \end{minipage}%
    \begin{minipage}{.5\textwidth}
     \centering
     \includegraphics[width=7.5cm, height=5cm]{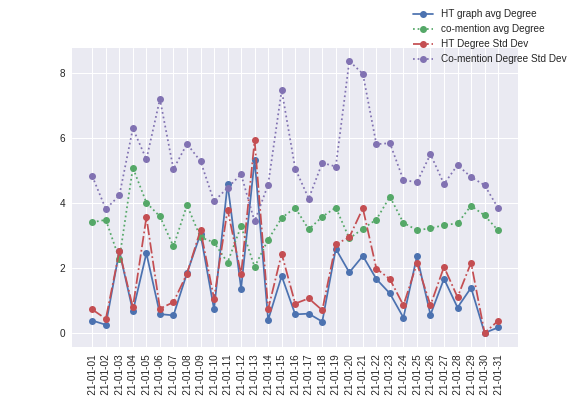}
      \caption{AVG Degree Distributions and Std Dev}
    \end{minipage}
    \caption{Various Metrics for Daily Partitioned Hashtag co-occurrences and User mentioned Graph from the Political Personalty Dataset. $Q_0$ retrieves tweets where Kamala Harris is mentioned in a hashtag, text body or user mention.}
    \label{fig:kamala}
\end{figure*}

\noindent  \textbf{Propagativeness Disparity:} The concept of propagativeness builds on the concept of navigability. Propagativeness attempts to capture how strongly the network is spreading information through a navigable subgraph $S$. We illustrate the concept with a network constructed over tweets where a political personality (Senator Kamala Harris in this example) is mentioned in January 2021. The three rows in Figure \ref{fig:kamala} show the network characteristics of the subregions of this graph, related, respectively to the themes of \texttt{\#ados} and ``black lives matter'' (Row 1), Captiol insurrection (Row 2) and Socioeconomic issues related to COVID-19 including stimulus funding ad business reopening (Row 3). In earlier work \citep{zheng2019social}, we have shown that a well known propagation mechanism for tweets is to add user-mentions to improve the reach of a message - hence the user-mention subgraph is indicative of propagative activity. In Figure \ref{fig:kamala}, we compare the hashtag activity (measured by the Hashtag subgraph) and the mention activity (size of the mention graph) in these three subgraphs. Figure \ref{fig:kamala} (e) shows a low and fairly steady size of the user mention activity in relation to the hashtag activity on the same topic, and these two indicators are not strongly correlated. Further, Figure \ref{fig:kamala} (f) shows that the mean and standard deviation of node degree of hashtag activity are fairly close, and the average degree of user co-mention (two users mentioned in the same tweet) graph is relatively steady over the period of observation -- showing low propagativeness. In contrast, Row 1 and Row 2 show sharper peaks. But the curve in Figure \ref{fig:kamala} (c) declines and has low, uncorrelated user mention activity. Hence, for this topic although there is a lot of discussion (leading to high navigability edges), the propagativeness is quite low. In comparison, Figure \ref{fig:kamala} (a) shows a strong peak and a stronger correlation be the two curves indicating more propagativeness. The higher standard deviation in the co-mention node degree over time (Figure \ref{fig:kamala} (b)) also shows the making of more propagation around this topic compared to the others.

We capture propagativeness using current flow betweenness centrality \citep{brandes2005centrality} which is based on Kirchoff's current laws. We combine this with the average neighbor degree of the nodes of $S$ to measure the spreading propensity of $S$. The current flow betweenness centrality is the portion of all-pairs shortest paths that pass through a node, and the average neighbor degree is the average degree of the neighborhood of each node. If a subgraph has higher current flow betweenness centrality plus a higher average neighbor degree, the network should have faster communicability.  Let $\alpha_{ij}$ be the number of shortest paths from node $i$ to $j$ and $\alpha_{ij}(n)$ is the number of paths passes through the node $n$. Then the current flow betweenness centrality: $$C_{nb}(n)= \sum_{(i,j)\in V} \frac{\alpha_{ij}(n)}{\alpha_{ij}}$$

Suppose the distribution of the current flow betweenness centrality of two subgraphs $A$ and $B$ is $p_1$ and $p_2$ respectively, and distribution of the reference graph is $p_t$. Also the distribution of the $\beta_{n}$, the average neighbor degree of the node $n$, for the subgraph $A$ and $B$ is $\gamma_1$ and $\gamma_1$ respectively, and the reference distribution is $\gamma_t$. If the condition
    $$\Delta(p_t, p_1) * \Delta(\gamma_t, \gamma_1) < \Delta(p_t, p_2) *  \Delta(\gamma_t, \gamma_2)$$
holds, we can conclude that subgraph $B$ is a faster propagating network than subgraph $A$. This measure is of interest in a social media based on the observation that misinformation/disinformation propagation groups either try to increase the average neighbor degree by adding fake nodes or try to involve influential nodes with high edge centrality to propagate the message faster \citep{besel2018full}.  

\medskip
\noindent    \textbf{Subgroups within a Candidate Subgraph:} 
    The purpose of the last metric is to determine whether a candidate subgraph identified using the previous measures need to be further decomposed into smaller subgraphs. We use subgraph centrality \citep{estrada2005subgraph} and coreness of nodes as our metrics. 
    The subgraph centrality measures the number of subgraphs a vertex participates in, and the core number of a node is the largest value $k$ of a $k$-core containing that node. So a subgraph for which the core number and subgraph centrality distributions are right-skewed compared to the background subgraph are (i) either split around high-coreness nodes, or (ii) reported to the user as a mixture of  diverse topics. 
The node grouping, per-group subgraph generation and candidate subgraph identification process is presented in Algorithm \ref{alg:graph-metrics}. In the algorithm, function \textit{cut2bin} extends the cut function, which compares the histograms of the two distributions whose domains (X-values) must overlap, and produces equi-width bins to ensure that two histograms (i.e., frequency distributions) have compatible bins.

\subsection{The Testing Process}
\label{sec:testing}

\begin{algorithm}
\scriptsize
\caption{Graph Construction Algorithm}
\label{alg:graph-metrics}
\SetKwProg{ComputeMetrics}{Function \emph{ComputeMetrics}}{}{end}
INPUT : $Q_{out}$ Output of the query, $L$ Graph construction rules, $gv$ grouping variable, $th_{size}$ is the minimum size of the subgraph\;
\SetKwProg{gmetrics}{Function \emph{gmetrics}}{}{end}
\SetKwProg{CompareHistograms}{Function \emph{CompareHistograms}}{}{end}
\gmetrics{($Q_{out}$, $L$, $groupVar$)}{
G[]$\leftarrow$ ConstructGraph($Q_{out}$, $L$)\;
$T \leftarrow$ []\;
\For{$g \in G $}{
     $t_{\alpha} \leftarrow$ ComputeMetrics(g)\; 
     $T.push(t_{alpha})$\;
    
    }
return $T$
}
\ComputeMetrics{(Graph g)}{
$m\leftarrow[]$\;
$m.push(eigenVectorCentrality(g))$\;
.........
$m.push(coreNumber(g))$\;
return $m$
}
\CompareHistograms{(List $t_{1}$, List $x_{2}$)}{
$s_g \leftarrow cut2bin(x_2, bin_{edges})$\;
$bin_{edges} \leftarrow$ getBinEdges($x_{2}$)\;
$t_g \leftarrow cut2bin(t_1, bin_{edges})$\;

$\beta_{js} \leftarrow distance.jensenShannon(t_g, s_g)$\;
$h_t \leftarrow histogram(t_g, s_g,bin_{edges} )$\;
return $\beta_{js}, h_t, bin_{edges}$\;
}
\end{algorithm}

\begin{algorithm}
 \scriptsize
\caption{Graph Discovery Algorithm}
\label{alg:discovery-algo}
\SetKwProg{discover}{Function \emph{discover}}{}{end}
\KwIn{ Set of all subgraphs divergence $\sigma$}
\KwOut{Feature vectors $v_1$, $v_2$, $v_3$, List for re-partition recommendations $l$}
$ev$ : eigenvector centrality\;
$ec$ : edge current flow betweenness centrality\;
$nc$ : current flow betweenness centrality\;
$\mu$ : core number\;
$z$ : average neighbor degree\;
\discover{($\sigma$)}{
\For{any two set of divergence from $\sigma_1$ ans $\sigma_2$}{
\If{$\sigma_2(ev) > \sigma_1(ev)$}{
 $v_1(\sigma_2) = v_1(\sigma_2) + 1$\;
 \If{$\sigma_2(ec) > \sigma_1(ec)$}{
 $v_2(\sigma_2) = v_2(\sigma_2) + 1$\;
 \If{($\sigma_2(nc)+ \sigma_2(\mu)) > (\sigma_1(ec) + \sigma_2(\mu)$)}{
 $v_3(\sigma_2) = v_3(\sigma_2) + 1$\;
  }
  \If{($\sigma_2(sc)+ \sigma_2(z)) > (\sigma_1(sc) + \sigma_2(z)$)}{
 $l(\sigma_2) = 1$\;
  }
 }
 }
 }
}
\end{algorithm} 
The discovery algorithm’s input is the list of divergence values of two candidate sets computed against the same reference graph. It produces four lists at the end. Each of the first three lists contains one specific factor of interestingness of the subgraph. The most interesting subgraph should present in all three vectors. 
If the subgraph has many cores and is sufficiently dense, then the system considers the subgraph to be \textit{uninterpretable} and sends it for re-partitioning. 
Therefore, the fourth list contains the subgraph that should partition again. Currently, our repartitioning strategy is to take subsets of the original keyword list provided by the user at the beginning of the discovery process to re-initiate the discovery process for the dense, uninterpretable subgraph.\\

\noindent The output of each metric produces a value for each participant node of  the input. However, to compare two different candidates, in terms of the metrics mentioned above,  we need to convert them to comparable histograms by applying a binning function depending on the data type of the grouping function. 
\\
\noindent \textit{Bin Formation (cut2bin):} Cut is a conventional operator (available with R, Matlab, Pandas etc. ) segments and sorts data values into bins. The cut2bin is an extension of a standard cut function, which compares the histograms of the two distributions whose domains (X-values) must overlap. The cut function accepts as input a set of set of node property values (e.g., the centrality metrics), and optionally a set of edge boundaries for the bins. It returns the histograms of distribution. Using the cut, first, we produce $n$ equi-width bins from the distribution with the narrower domain. Then we extract bin edges from the result and use it as the input bin edges to create the wider distribution`s cut.  This enforces the histograms to be compatible. In case one of the distribution is known to be a reference distribution (distribution from the background graph) against which the second distribution is compared, we use the reference distribution for equi-width binning and bin the second distribution relative to the first.
\\
\noindent The $CompareHistograms$ function uses the \textit{cut2bin} function to produce the histograms, and then computes the JS Divergence on the comparable histograms. The $CompareHistograms$ function returns the set of divergence values for each metric of a subgraph, which is the input of the discovery algorithm. The function requires the user to specify which of the compared graphs should be considered as a reference -- this is required to ensure that our method is scalable for large background graphs (which are typically much larger than the interesting subgraphs). If the background graph is very large, we take several random subgraphs from this graph to ensure they are representative before the actual comparisons are conducted. To this end, we adopt the well-known random walk strategy.
 
\noindent In the algorithm $v_1$, $v_2$ and $v_3$ are the three vectors to store the interestingness factors of the subgraphs, and $l$ is the list for repartitioning.  For two subgraphs, if one of them qualified for $v_1$ means, the subgraph contains higher centrality than the other. In that case, it increases the value of that qualified bit in the vector by one. Similarly, it increases the value of  $v_2$ by one,  if the same candidate has high navigability. Finally, it increases the $v_3$, if it has higher propagativeness.  The algorithm selects the top-$k$ scores of candidates from each vector, and marks them interesting.


\begin{table*}[]
\caption{Dataset Descriptions }
\label{tab:dataset}
\begin{tabular}{|l|l|l|l|l|l|l|l|l|l|}
\hline
\textbf{Data Set} & \textbf{\begin{tabular}[c]{@{}l@{}}Total \\ Collection\\  Size\end{tabular}} & \textit{\textbf{\begin{tabular}[c]{@{}l@{}}Sub\\  Query\end{tabular}}} & \textbf{\begin{tabular}[c]{@{}l@{}}Network \\ Type\end{tabular}} & \textbf{\begin{tabular}[c]{@{}l@{}}Total \\ Tweets\end{tabular}} & \textbf{\begin{tabular}[c]{@{}l@{}}Unique \\ nodes\end{tabular}} & \textbf{\begin{tabular}[c]{@{}l@{}}Unique \\ edges\end{tabular}} & \textbf{\begin{tabular}[c]{@{}l@{}}Self\\ Loop\end{tabular}} & \textbf{Density} & \textbf{\begin{tabular}[c]{@{}l@{}}Avg\\ Degree\end{tabular}} \\ \hline
\textit{\begin{tabular}[c]{@{}l@{}}Kamala \\ Harris\end{tabular}} & 12469480 & \textit{\begin{tabular}[c]{@{}l@{}}Capitol \\ Attack\end{tabular}} & \begin{tabular}[c]{@{}l@{}}Hashtag \\ co-occur\end{tabular} & 164397 & 1398 & 7801 & 16 & 0.0025 & 4.3 \\ \hline
\textit{} &  & \textit{} & \begin{tabular}[c]{@{}l@{}}user\\ co-mention\end{tabular} & 164397 & 8012 & 48604 & 87 & 0.00012 & 3.19 \\ \hline
\textit{} &  & \textit{\#ADOS} & \begin{tabular}[c]{@{}l@{}}Hashtag \\ co-occur\end{tabular} & 158419 & 3671 & 10738 & 29 & 0.0015 & 5.8 \\ \hline
\textit{} &  & \textit{} & \begin{tabular}[c]{@{}l@{}}user \\ co-mention\end{tabular} & 158419 & 30829 & 39865 & 49 & 8.3 & 2.5 \\ \hline
\textit{} &  & \textit{\begin{tabular}[c]{@{}l@{}}Economic \\ Issues\end{tabular}} & \begin{tabular}[c]{@{}l@{}}Hashtag \\ co-occur\end{tabular} & 36678 & 1278 & 1828 & 4 & 0.0022 & 2.8 \\ \hline
\textit{} &  & \textit{} & \begin{tabular}[c]{@{}l@{}}user \\ co-mention\end{tabular} & 36678 & 6971 & 11584 & 19 & 0.0004 & 3.4 \\ \hline
\textit{Joe Biden} & 45258151 & \textit{\begin{tabular}[c]{@{}l@{}}Capitol \\ Attack\end{tabular}} & \begin{tabular}[c]{@{}l@{}}Hashtag \\ co-occur\end{tabular} & 676898 & 7728 & 21422 & 50 & 0.00071 & 5.49 \\ \hline
\textit{} &  & \textit{} & \begin{tabular}[c]{@{}l@{}}user \\ co-mention\end{tabular} & 676898 & 82046 & 101646 & 130 & 3.0146 & 2.473 \\ \hline
\textit{} &  & \textit{\#ADOS} & \begin{tabular}[c]{@{}l@{}}Hashtag \\ co-occur\end{tabular} & 183765 & 3007 & 11008 & 29 & 0.002 & 3.85 \\ \hline
\textit{} &  & \textit{} & \begin{tabular}[c]{@{}l@{}}user \\ co-mention\end{tabular} & 158419 & 29547 & 40932 & 56 & 9.3 & 2.7 \\ \hline
\textit{} &  & \textit{\begin{tabular}[c]{@{}l@{}}Economic \\ Issues\end{tabular}} & \begin{tabular}[c]{@{}l@{}}Hashtag \\ co-occur\end{tabular} & 138754 & 2961 & 5733 & 10 & 0.0013 & 3.87 \\ \hline
\textit{} &  & \textit{} & \begin{tabular}[c]{@{}l@{}}user \\ co-mention\end{tabular} & 138754 & 21417 & 19691 & 23 & 8.5 & 1.83 \\ \hline
\textit{Vaccine} & 24172676 & \textit{\begin{tabular}[c]{@{}l@{}}Vaccine\\ Anti-vax\end{tabular}} & \begin{tabular}[c]{@{}l@{}}Hashtag \\ co-occur\end{tabular} & 1000000 & 18809 & 24195 & 44 & 2.52 & 2.5 \\ \hline
\textit{} &  & \textit{} & \begin{tabular}[c]{@{}l@{}}user \\ co-mention\end{tabular} & 1000000 & 203211 & 41877 & 46 & 2.02 & 0.4 \\ \hline
\textit{} &  & \textit{Covid Test} & \begin{tabular}[c]{@{}l@{}}Hashtag \\ co-occur\end{tabular} & 1000000 & 26671 & 45378 & 69 & 0.00012 & 3.4 \\ \hline
\textit{} &  & \textit{} & \begin{tabular}[c]{@{}l@{}}user \\ co-mention\end{tabular} & 1000000 & 188761 & 83656 & 109 & 4.67 & 0.886 \\ \hline
 &  & economy & \begin{tabular}[c]{@{}l@{}}hashtag \\ co-occur\end{tabular} & 917890 & 3002 & 4395 & 9 & 0.0009 & 2.9 \\ \hline
 &  &  & \begin{tabular}[c]{@{}l@{}}user \\ co-mention\end{tabular} & 917890 & 20590 & 8528 & 13 & 4.023 & 0.8 \\ \hline
\end{tabular}
\end{table*}

\section{Experiments}
\label{sec:experiments}
\subsection{Data Sets}
\label{sec:datasets}
Data sets used for the experiments are tweets collected using the Twitter Streaming API using a set of domain-specific, hand-curated keywords. We used three data sets, all collected between the 1st and the 31st of January, 2021. The first two sets are for political personalities. The Kamala Harris data set was collected by using variants of Kamala Harris's name together with her Twitter handle and hashtags constructed from her name. The second data set was similarly constructed for Joe Biden. The third data set was collected during the COVID-19 pandemic. The keywords were selected based on popularly co-occurring terms from Google Trends. We selected the terms manually to ensure that they are related to the pandemic and vaccine related issues (and not, for example, partisan politics). Table \ref{tab:dataset} presents a quantitative summary on the three data sets. We used a set of subqueries to find subsets from each of our datasets. These subqueries are temporal, and represent trending terms that stand for the emerging issues. For the first two datasets, we used three themes to construct the background graphs: a) capitol attack and insurrection, b) Black Lives Matter and ADOS movement, and c) American economic crisis. and recovery efforts. For the vaccine data set, we selected two subsets of posts, one for vaccine-related concerns, anti-vaccine movements, and related issues, and the second for covid testing and infection-related issues. The vaccine data set is larger, and the content is more diverse than the first two. 
All the data sets are publicly available from The Awesome Lab \footnote{https://code.awesome.sdsc.edu/awsomelabpublic/datasets/int-springer-snam/)}

In the experiments, we constructed two subgraphs from each subquery. The first is the “hashtag co-occurrence” graph, where each hashtag is a node, and  they are connected through an edge if they coexist in a tweet. The second is the “user co-mention” graph, where each user is a node, and  there is an edge between two nodes if a tweet mentioned them jointly. Intuitively, the hashtag co-occurrence subgraph captures topic prevalence and propagation, whereas the co-mention subgraph captures the tendency to influence and propagate messages to a larger audience. Our goal is to discover surprises (and lack thereof) in these two aspects for our data sets.

We note that the dataset chosen is from a month where the US experienced a major event in the form of the Capitol Attack, and a new administration was sworn in. This explains why the number of tweets in ``Capitol Attack'' subgraph is high for both politicians in this week, and not surprisingly it is also the most discussed topic as evidenced by the high average node degree. Therefore, this ``selection bias'' sets our expectation for subjective interestingness -- given the specific week we have chosen, this issue will dominate most social media conversations in the USA. We also observe the low ratio of the number of unique nodes to the number of tweets, signifying the high number of retweets, that signals a form of information propagation over the network. The propagativeness of the network during this eventful week is also evidenced by the fact that the unique node count of a co-mention network is almost 75\% - 88\% higher on average compared to the hashtag co-occur network of the same class. In Section \ref{sec:analysis}, we show how our interestingness technique performs in the face of this dataset.


\begin{figure*}[t] 
    \begin{minipage}{.33\textwidth}
      \includegraphics[width=6.5cm, height=5.5cm]{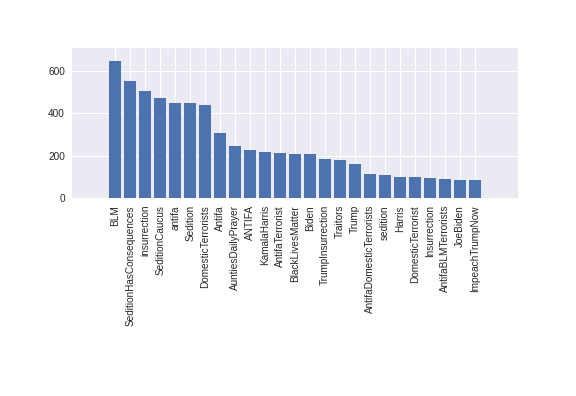}
        \caption{Capitol Attack}
           \label{fig:hist-results-harris-capitol}
    \end{minipage}%
    \begin{minipage}{.33\textwidth}
      \includegraphics[width=6.5cm, height=5.5cm]{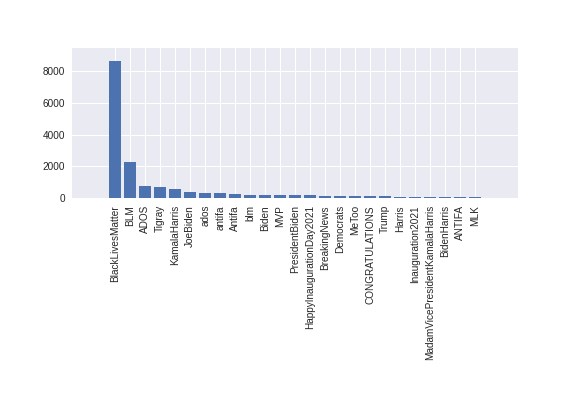}
        \caption{\#ADOS}
           \label{fig:hist-results-harris-ados}
    \end{minipage}%
    \begin{minipage}{.33\textwidth}
      \includegraphics[width=6.5cm, height=5.5cm]{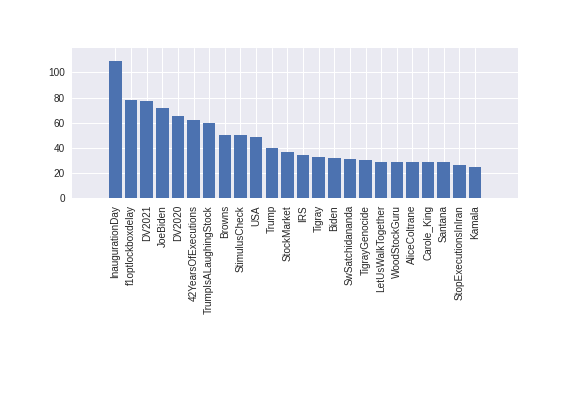}
        \caption{Economic issues}
           \label{fig:hist-results-harris}
    \end{minipage}%

     \caption{Top Hashtag Distributions of the Kamala Harris Data set }
     \label{fig:hist-results-harris}
\end{figure*}
\begin{figure*}[t] 
    \begin{minipage}{.33\textwidth}
      \includegraphics[width=6.5cm, height=5.5cm]{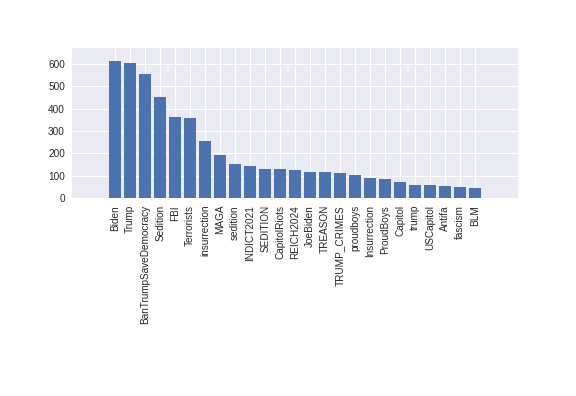}
        \caption{Capitol Attack}
           \label{fig:hist-results-biden-capitol}
    \end{minipage}%
    \begin{minipage}{.33\textwidth}
      \includegraphics[width=6.5cm, height=5.5cm]{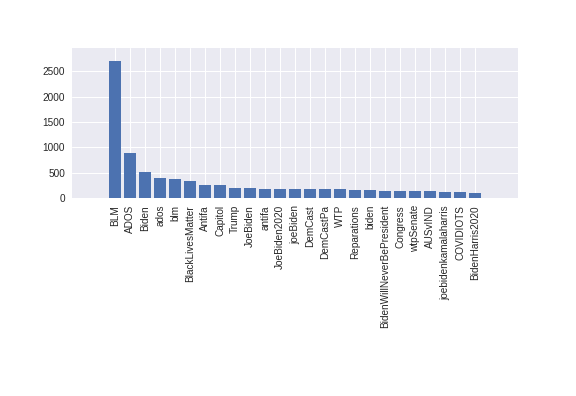}
        \caption{\#ADOS Issues}
           \label{fig:hist-results-biden-ados}
    \end{minipage}%
    \begin{minipage}{.33\textwidth}
      \includegraphics[width=6.5cm, height=5.5cm]{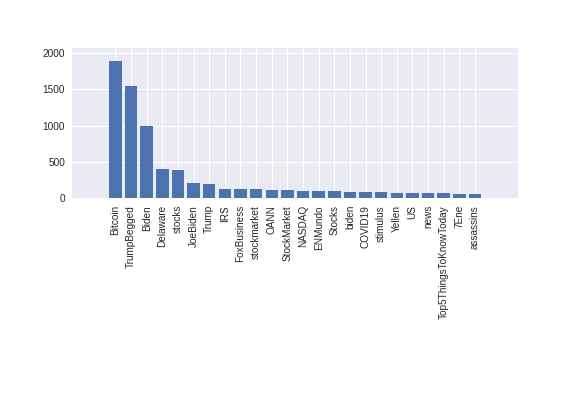}
        \caption{Economic issues}
           \label{fig:hist-results-biden-eco}
    \end{minipage}%

     \caption{Top Hashtag Distributions of the Joe Biden set }
     \label{fig:hist-results-biden}
\end{figure*} 
\begin{figure*}[t]
    \begin{minipage}{.33\textwidth}
      \includegraphics[width=6.5cm, height=5.5cm]{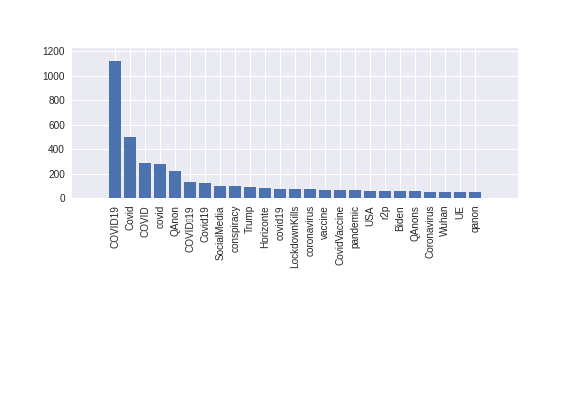}
        \caption{Vaccine anti-vaccine Issues}
           \label{fig:hist-results-vac-vac-anti-vax}
    \end{minipage}%
    \begin{minipage}{.33\textwidth}
      \includegraphics[width=6.5cm, height=5.5cm]{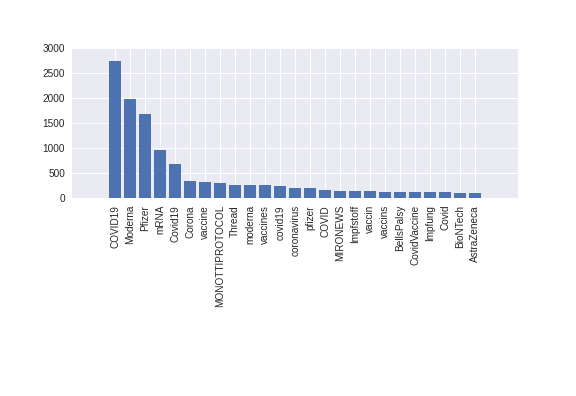}
        \caption{COVID-19 Test }
           \label{fig:hist-results-vac-covid}
    \end{minipage}%
    \begin{minipage}{.33\textwidth}
      \includegraphics[width=6.5cm, height=5.5cm]{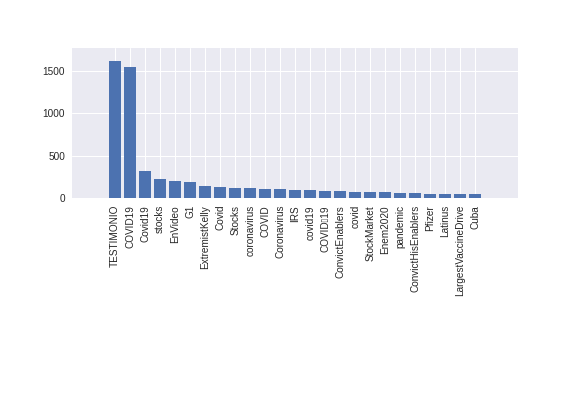}
        \caption{Economic issues}
           \label{fig:hist-results-vac-eco}
    \end{minipage}%
     \caption{Top Hashtag Distributions of the Vaccine Data set }
     \label{fig:hist-results-vac}
\end{figure*} 
\subsection{Experimental Setup}
\label{sec:setup}
The experimental setup has three steps a) data collection and archival storage, b)Indexing and storing data, and c) executing analytical pipelines. We used The AWESOME project’s continuous tweet ingestion system that collects all the tweets through Twitter 1\% REST API using a set of hand-picked keywords. We used the AWESOME Polysotre for indexing, storing, and search the data. For computation, we used the Nautilus facility of the Pacific Research Platform (PRP). Our hardware configurations are as follows. The Awesome server has 64 GB memory and 32 core processors, the Nautilus has 32 core, and 64 GB nodes. The data ingestion process required a large memory. Depending on the density of the data, this requirement varies. Similarly, centrality computation is a CPU bounded process. Performance optimizations that we implemented are outside the scope of this paper.   
\medskip
\begin{figure*}[t]
    \centering
    \begin{minipage}{.5\textwidth}
    \centering
      \includegraphics[width=7.5cm, height=5.5cm]{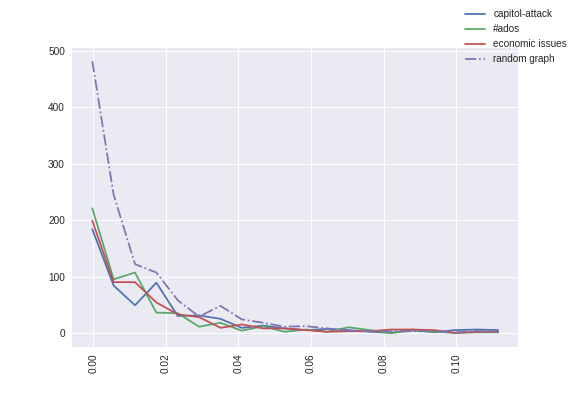}

        \caption{Eigenvector Centrality Disparity Hashtag Network}
          \label{fig:result-pipline-harris-ht-egc}
    \end{minipage}%
    \begin{minipage}{.5\textwidth}
     \centering
     \includegraphics[width=7.5cm, height=5.5cm]{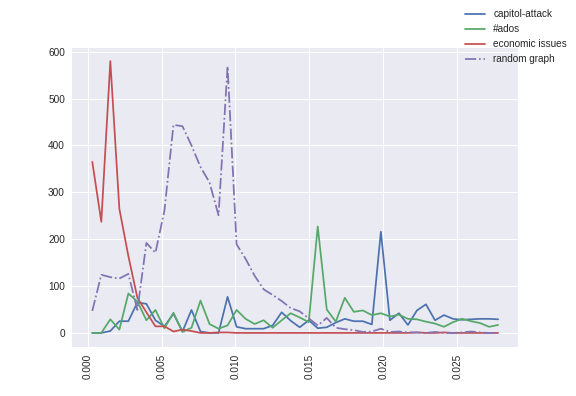}
       \caption{Topical Navigability Disparity Hashtag Network}
        \label{fig:result-pipline-harris-ht-nav}
    \end{minipage}
    \begin{minipage}{.5\textwidth}
    \centering
      \includegraphics[width=7.5cm, height=5.5cm]{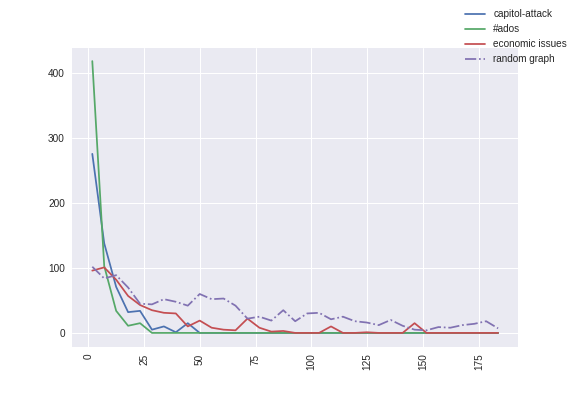}

        \caption{Propagativeness  Disparity Hashtag Network }
           \label{fig:result-pipline-harris-ht-prop}
    \end{minipage}%
    \begin{minipage}{.5\textwidth}
     \centering
     \includegraphics[width=7.5cm, height=5.5cm]{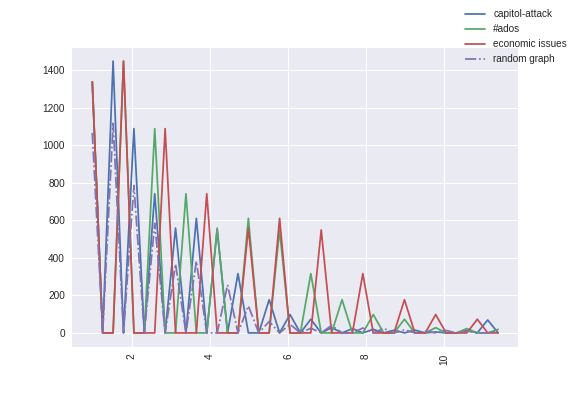}
       \caption{Eigenvector Centrality Disparity co-mention Network}
    \end{minipage}
    \begin{minipage}{.5\textwidth}
    \centering
      \includegraphics[width=7.5cm, height=5.5cm]{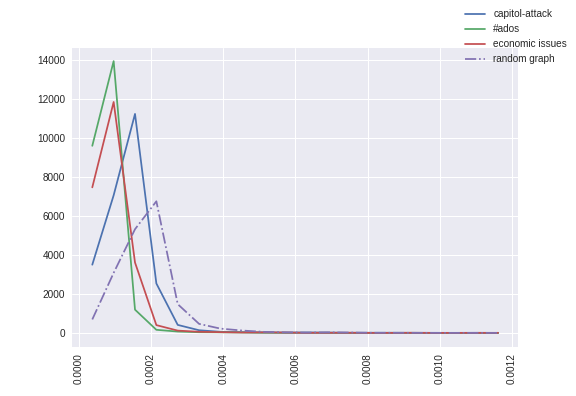}

        \caption{Topical Navigability Disparity co-mention Network}
           \label{fig:result-pipline-harris-co-mention-nav}
    \end{minipage}%
    \begin{minipage}{.5\textwidth}
     \centering
     \includegraphics[width=7.5cm, height=5.5cm]{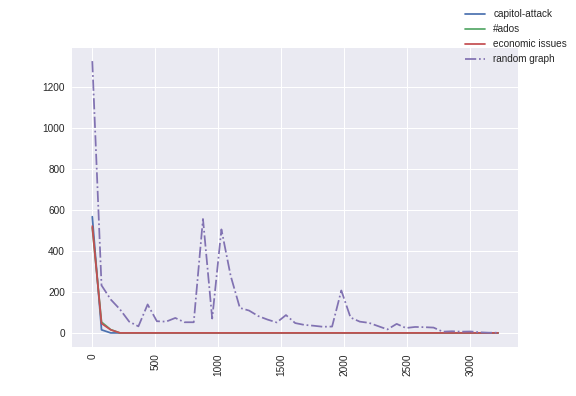}
       \caption{Propagativeness  Disparity co-mention Network}
        \label{fig:result-pipline-harris-co-mention-prop}
    \end{minipage}
    \caption{Comparative studies of all sub-queries using the "Kamala Harris" data set.}
     \label{fig:result-pipline-harris}
\end{figure*}
\begin{figure*}[t]
    \centering
    \begin{minipage}{.5\textwidth}
    \centering
      \includegraphics[width=7.5cm, height=5.5cm]{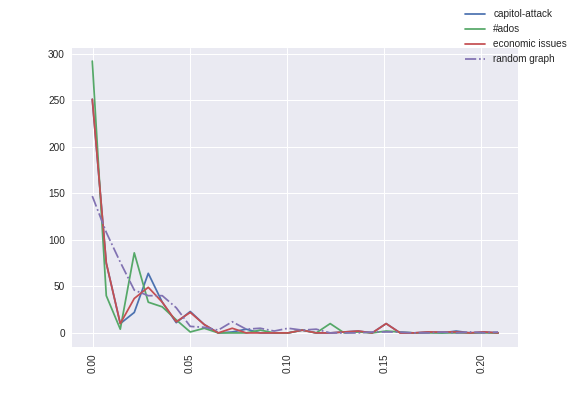}

        \caption{Eigenvector Centrality Disparity Hashtag Network}
           \label{fig:result-pipline-biden-ht-egc}
    \end{minipage}%
    \begin{minipage}{.5\textwidth}
     \centering
     \includegraphics[width=7.5cm, height=5.5cm]{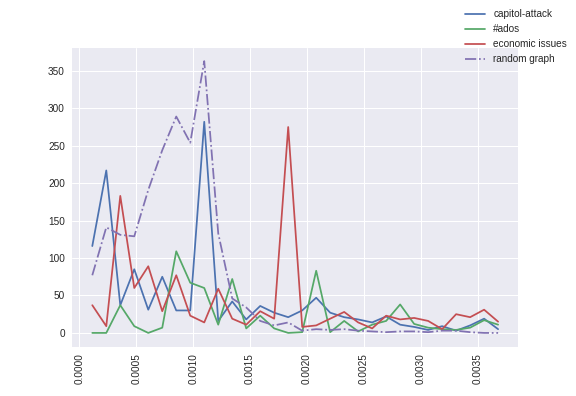}
       \caption{Topical Navigability Disparity Hashtag Network}
        \label{fig:result-pipline-biden-ht-nav}
    \end{minipage}
    \begin{minipage}{.5\textwidth}
    \centering
      \includegraphics[width=7.5cm, height=5.5cm]{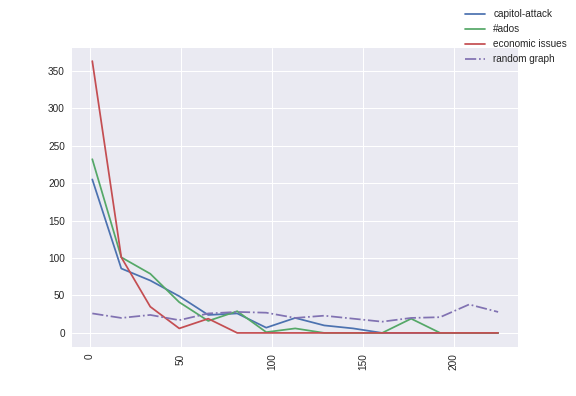}

        \caption{Propagativeness  Disparity Hashtag Network }
           \label{fig:result-pipline-biden-ht-prop}
    \end{minipage}%
    \begin{minipage}{.5\textwidth}
     \centering
     \includegraphics[width=7.5cm, height=5.5cm]{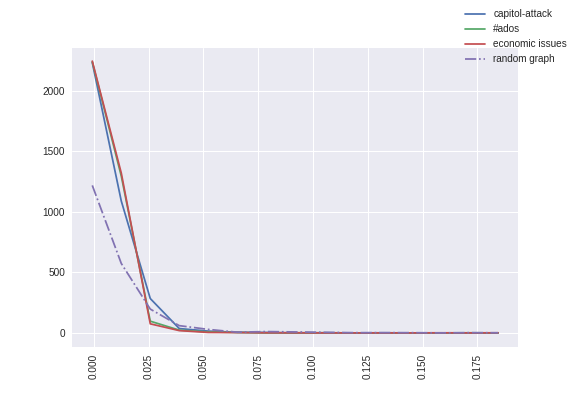}
       \caption{Eigenvector Centrality Disparity co-mention Network}
       \label{fig:result-pipline-biden-mention-egc}
    \end{minipage}
    \begin{minipage}{.5\textwidth}
    \centering
      \includegraphics[width=7.5cm, height=5.5cm]{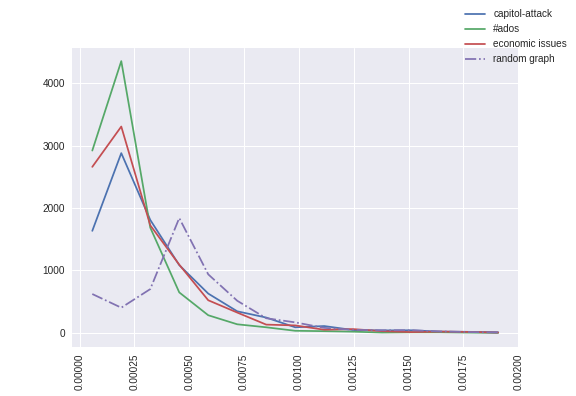}

        \caption{Topical Navigability Disparity co-mention Network}
           \label{fig:result-pipline-biden-mention-nav}
    \end{minipage}%
    \begin{minipage}{.5\textwidth}
     \centering
     \includegraphics[width=7.5cm, height=5.5cm]{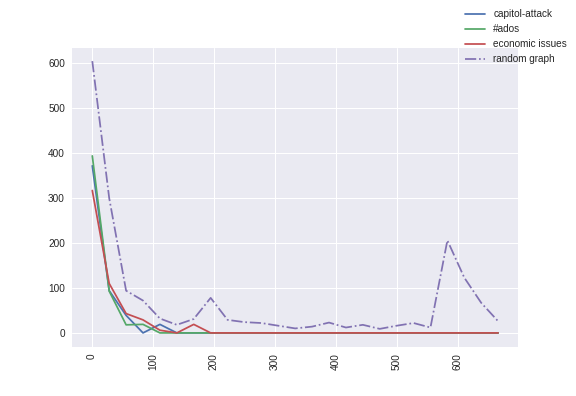}
       \caption{Propagativeness  Disparity co-mention Network}
      \label{fig:result-pipline-biden-mention-prop}
    \end{minipage}
   \caption{Comparative studies of all sub-queries using the "Joe Biden" data set.}
     \label{fig:result-pipline-biden}
\end{figure*}
\begin{figure*}
    \centering
    \begin{minipage}{.5\textwidth}
    \centering
      \includegraphics[width=7.5cm, height=5.5cm]{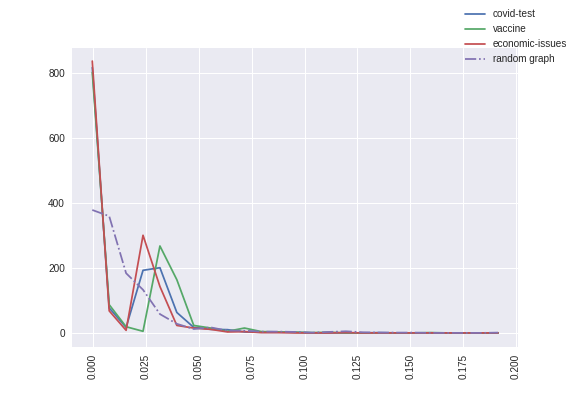}

        \caption{Eigenvector Centrality Disparity: "Vaccine" Hashtag Network}
           \label{fig:result-pipline-ht-egc}
    \end{minipage}%
    \begin{minipage}{.5\textwidth}
     \centering
     \includegraphics[width=7.5cm, height=5.5cm]{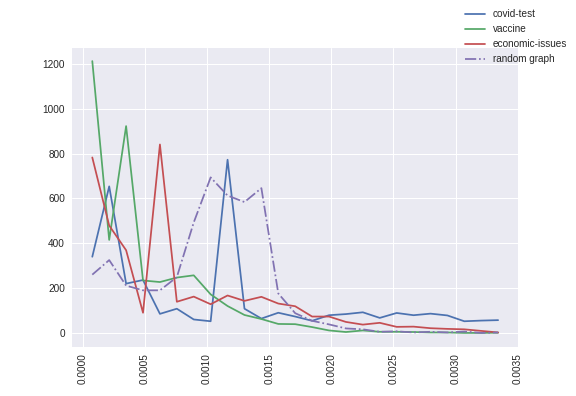}
       \caption{Topical Navigability Disparity: Vaccine Hashtag Network}
    \end{minipage}
    \begin{minipage}{.5\textwidth}
    \centering
      \includegraphics[width=7.5cm, height=5.5cm]{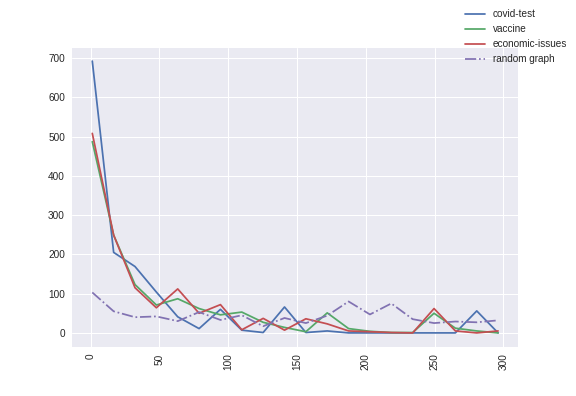}
        \caption{Propagativeness Disparity: "Vaccine" Hashtag Network }
           \label{fig:result-pipline-ht-vax}
    \end{minipage}%
    \begin{minipage}{.5\textwidth}
     \centering
     \includegraphics[width=7.5cm, height=5.5cm]{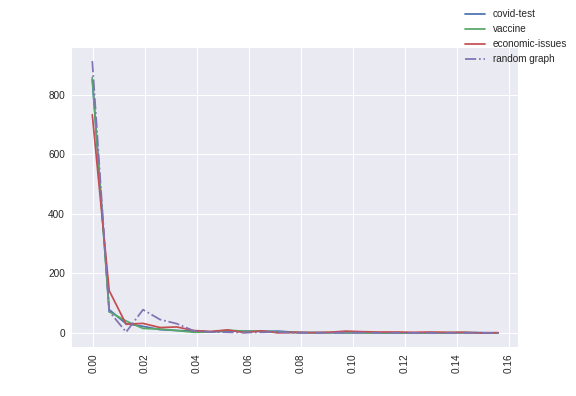}
       \caption{Eigenvector Centrality Disparity: "Vaccine" co-mention Network}
        \label{fig:result-pipline-mention-egc}
    \end{minipage}
    \begin{minipage}{.5\textwidth}
    \centering
      \includegraphics[width=7.5cm, height=5.5cm]{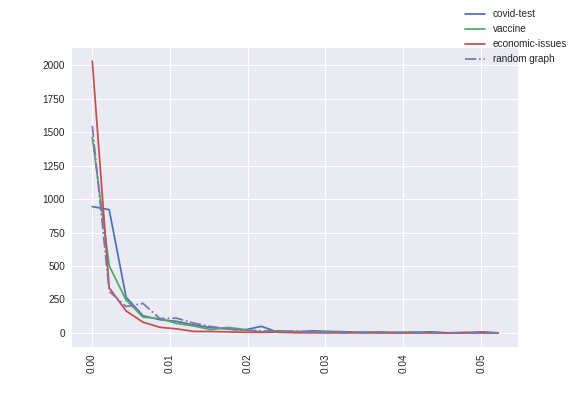}

        \caption{Topical Navigability Disparity: "Vaccine" co-mention Network}
           \label{fig:result-pipline-mention-nav}
    \end{minipage}%
    \begin{minipage}{.5\textwidth}
     \centering
     \includegraphics[width=7.5cm, height=5.5cm]{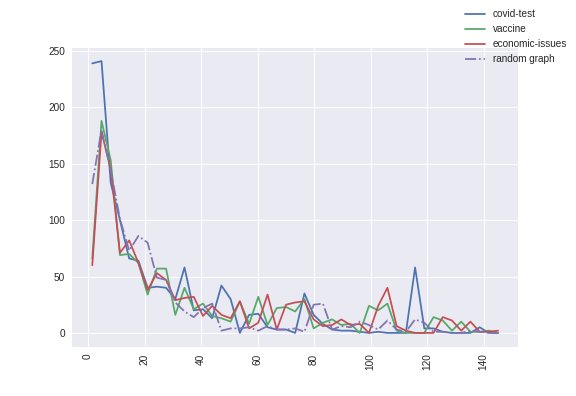}
       \caption{Propagativeness Disparity: "Vaccine" co-mention Network}
       \label{fig:result-pipline-mention-prop}
    \end{minipage}
    \caption{Comparative studies of all sub-queries using the "Vaccine" data set.}
     \label{fig:result-pipline-vax}
\end{figure*}
\begin{figure*}
    \begin{minipage}{.25\textwidth}
      \includegraphics[width=5cm, height=3.5cm]{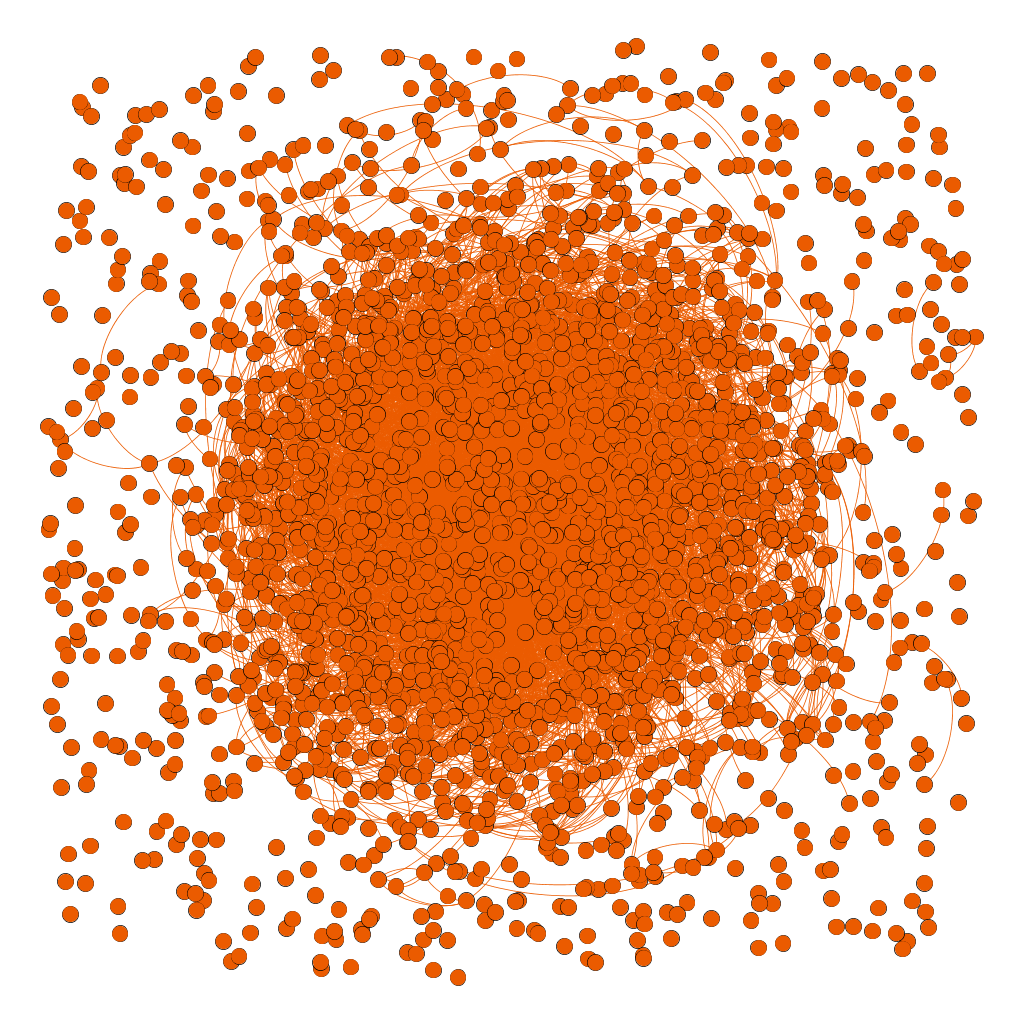}
        \caption{\#ADOS}
           \label{fig:core-peri-kamal-ados}
    \end{minipage}%
    \begin{minipage}{.25\textwidth}
      \includegraphics[width=5cm, height=3.5cm]{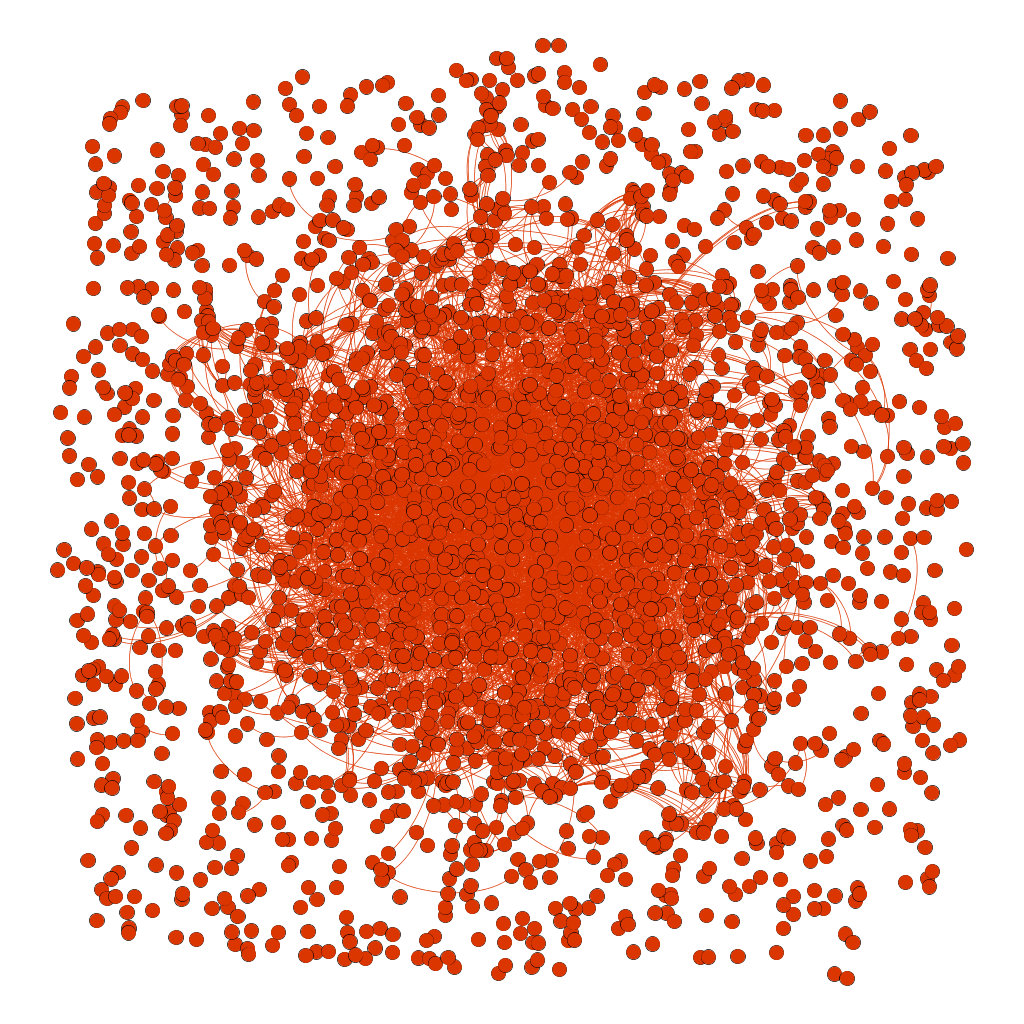}
        \caption{Capitol attack}
           \label{fig:core-peri-kamal-cap}
    \end{minipage}%
    \begin{minipage}{.25\textwidth}
      \includegraphics[width=5cm, height=3.5cm]{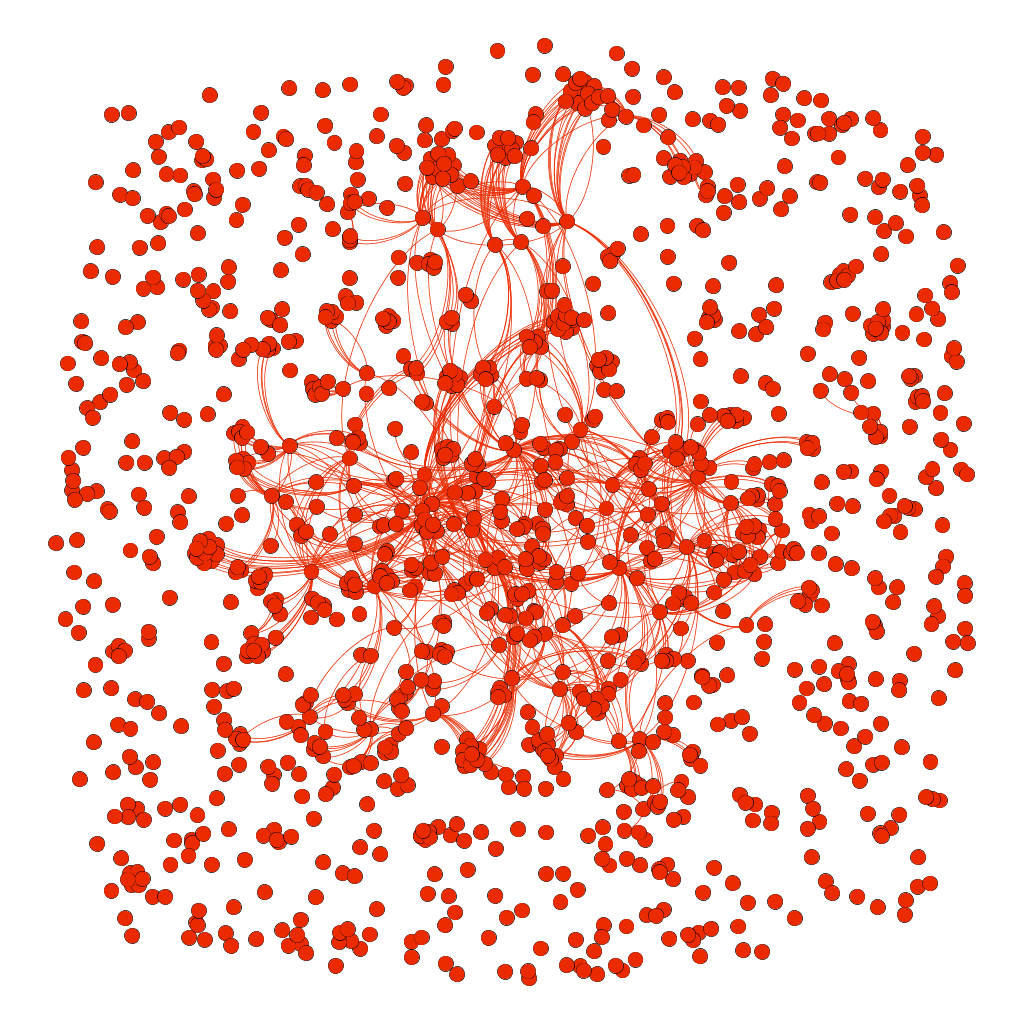}
        \caption{Economic issues}
           \label{fig:core-peri-kamal-eco}
    \end{minipage}%
    \begin{minipage}{.25\textwidth}
      \includegraphics[width=5cm, height=3.5cm]{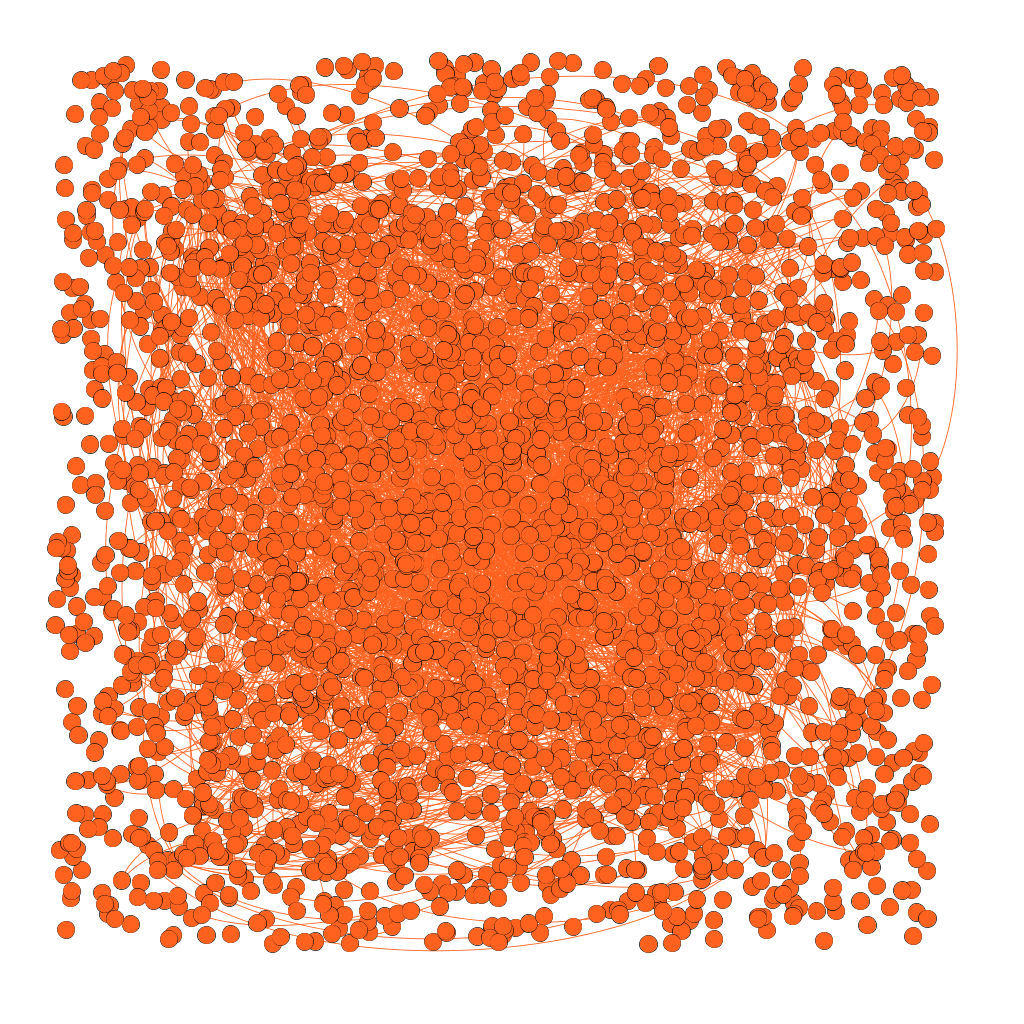}
        \caption{Random Graph}
           \label{fig:core-peri-kamal-random}
    \end{minipage}%
    \caption{Core and Periphery visualization of "Kamala Harris" Data set}
    \label{fig:core-peri-kamal}
\end{figure*}    
\begin{figure*}
    \begin{minipage}{.25\textwidth}
      \includegraphics[width=5cm, height=3.5cm]{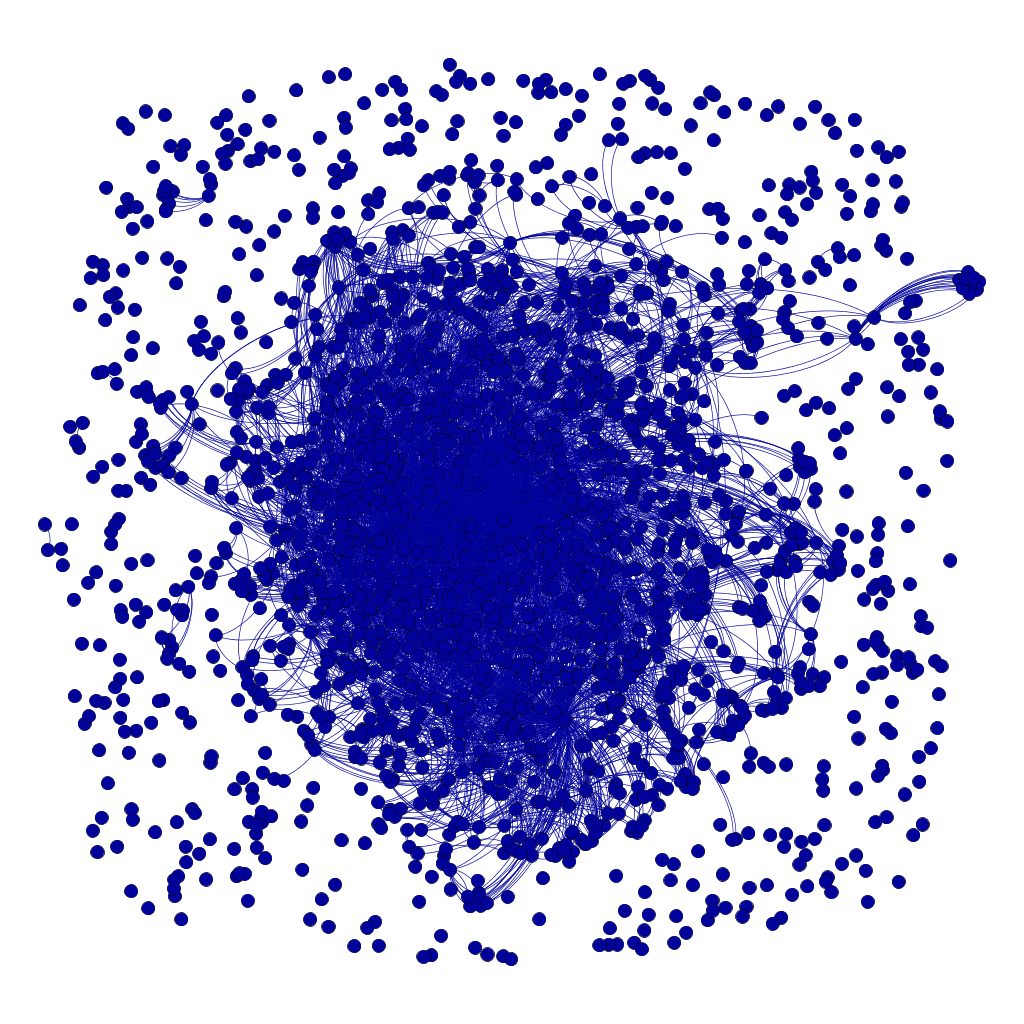}
        \caption{\#ADOS}
           \label{fig:core-peri-biden-ados}
    \end{minipage}%
    \begin{minipage}{.25\textwidth}
      \includegraphics[width=5cm, height=3.5cm]{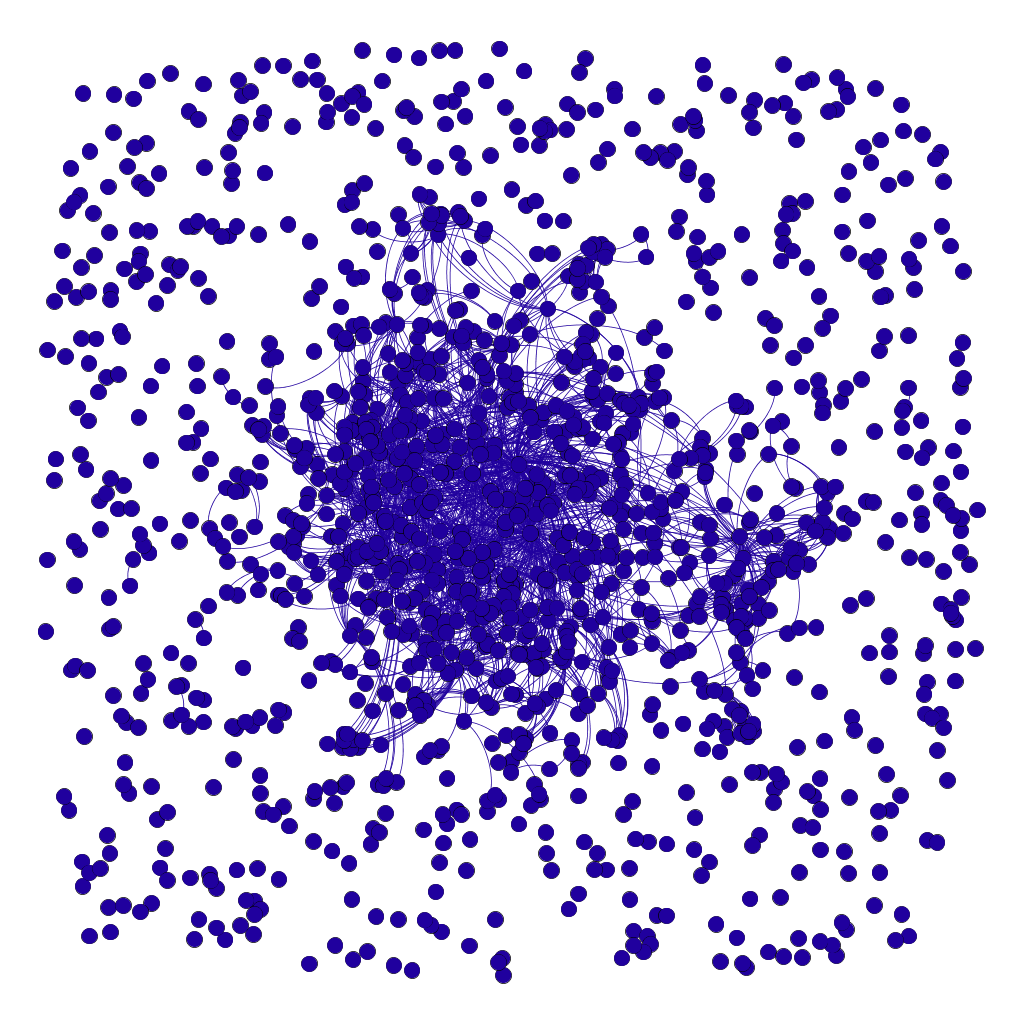}
        \caption{Capitol attack}
           \label{fig:core-peri-biden-capitol}
    \end{minipage}%
    \begin{minipage}{.25\textwidth}
      \includegraphics[width=5cm, height=3.5cm]{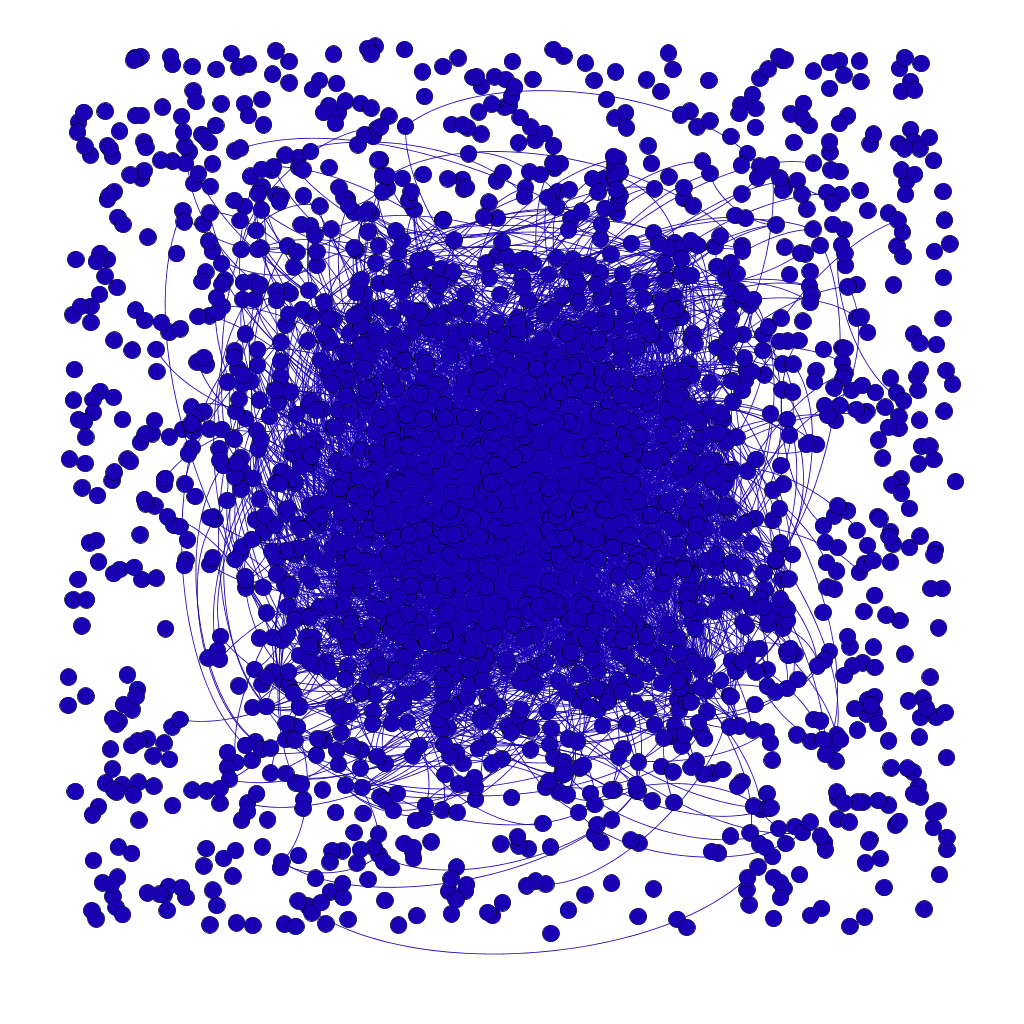}
        \caption{Economic issues}
            \label{fig:core-peri-biden-eco}
    \end{minipage}%
    \begin{minipage}{.25\textwidth}
      \includegraphics[width=5cm, height=3.5cm]{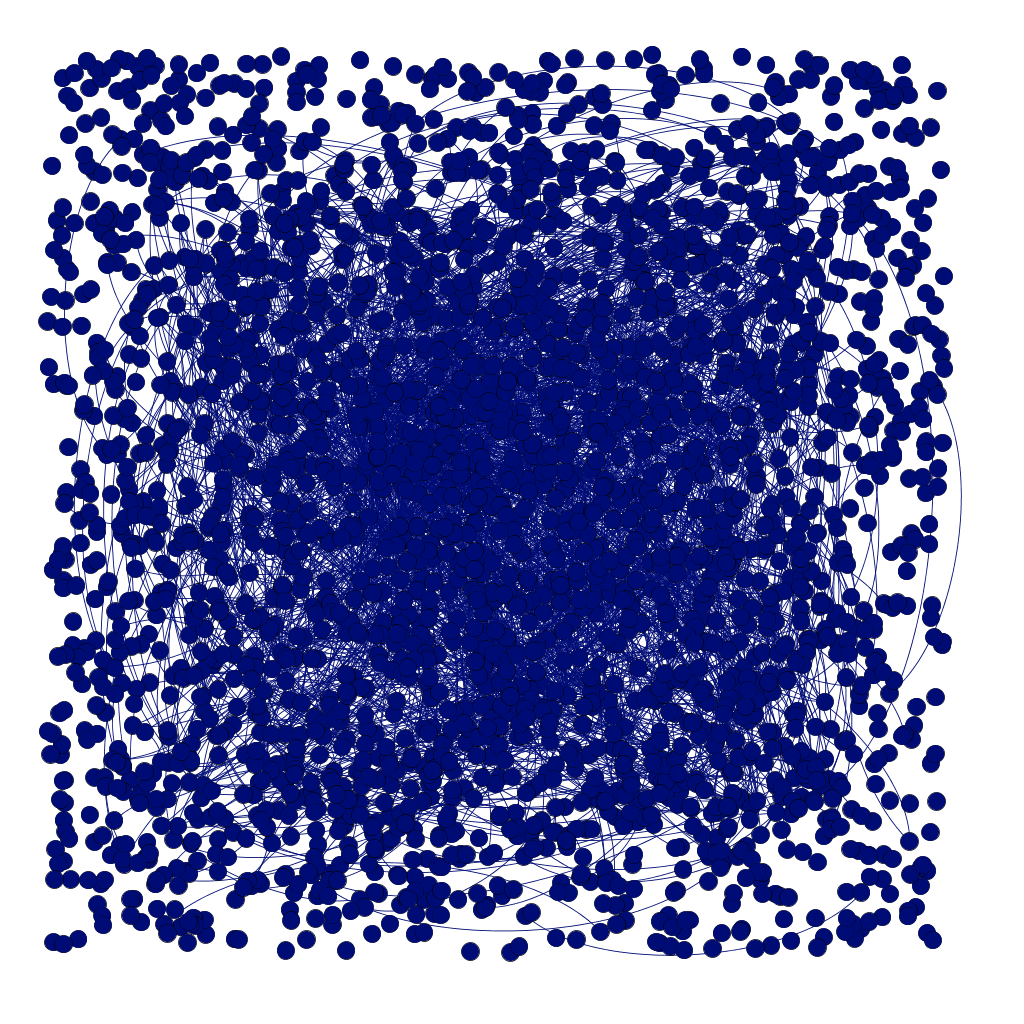}
        \caption{Random Data}
           \label{fig:core-peri-biden-random}
    \end{minipage}%
     \caption{Core and Periphery visualization of "Joe Biden" Data set}
      \label{fig:core-peri-biden}
\end{figure*}    
\begin{figure*}
    \begin{minipage}{.25\textwidth}
      \includegraphics[width=5cm, height=3.5cm]{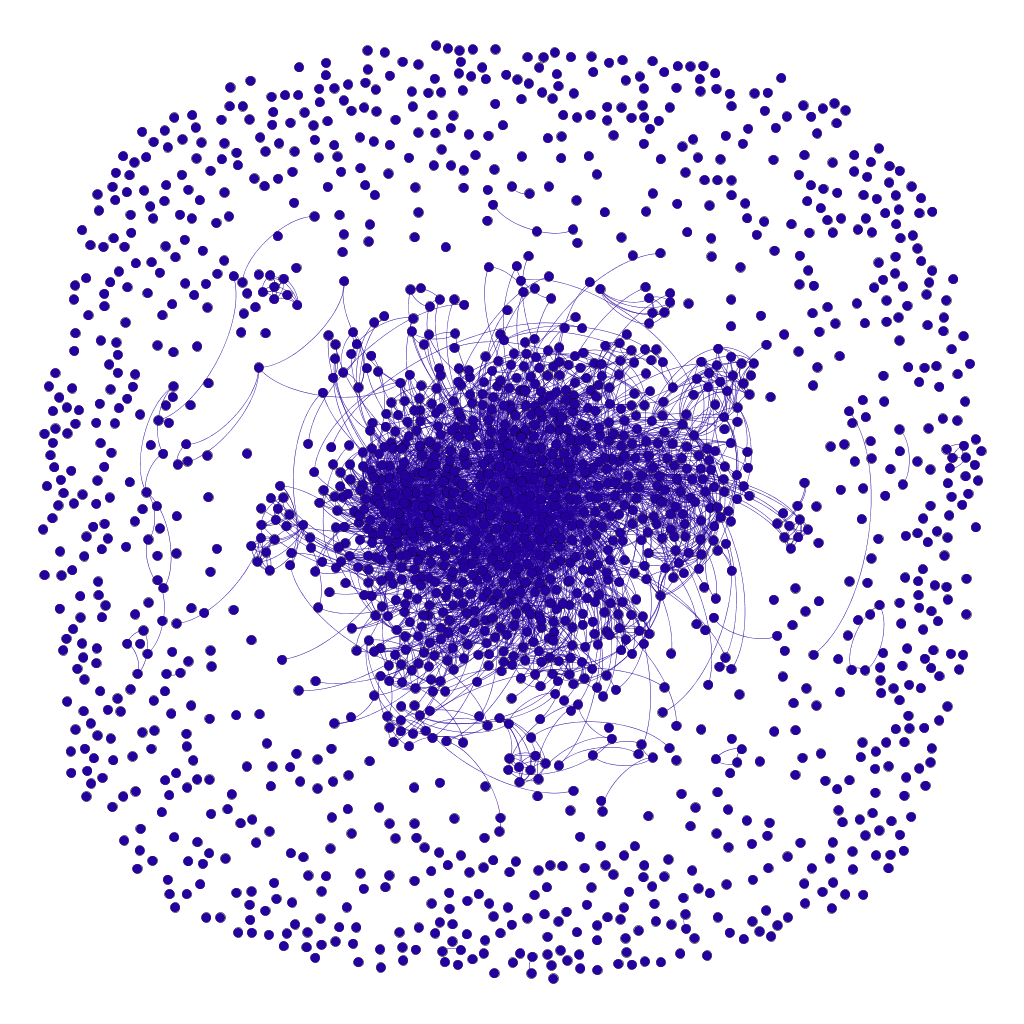}
        \caption{Vaccine Issues}
           \label{fig:core-peri-vax-vax-anti-vax}
    \end{minipage}%
    \begin{minipage}{.25\textwidth}
      \includegraphics[width=5cm, height=3.5cm]{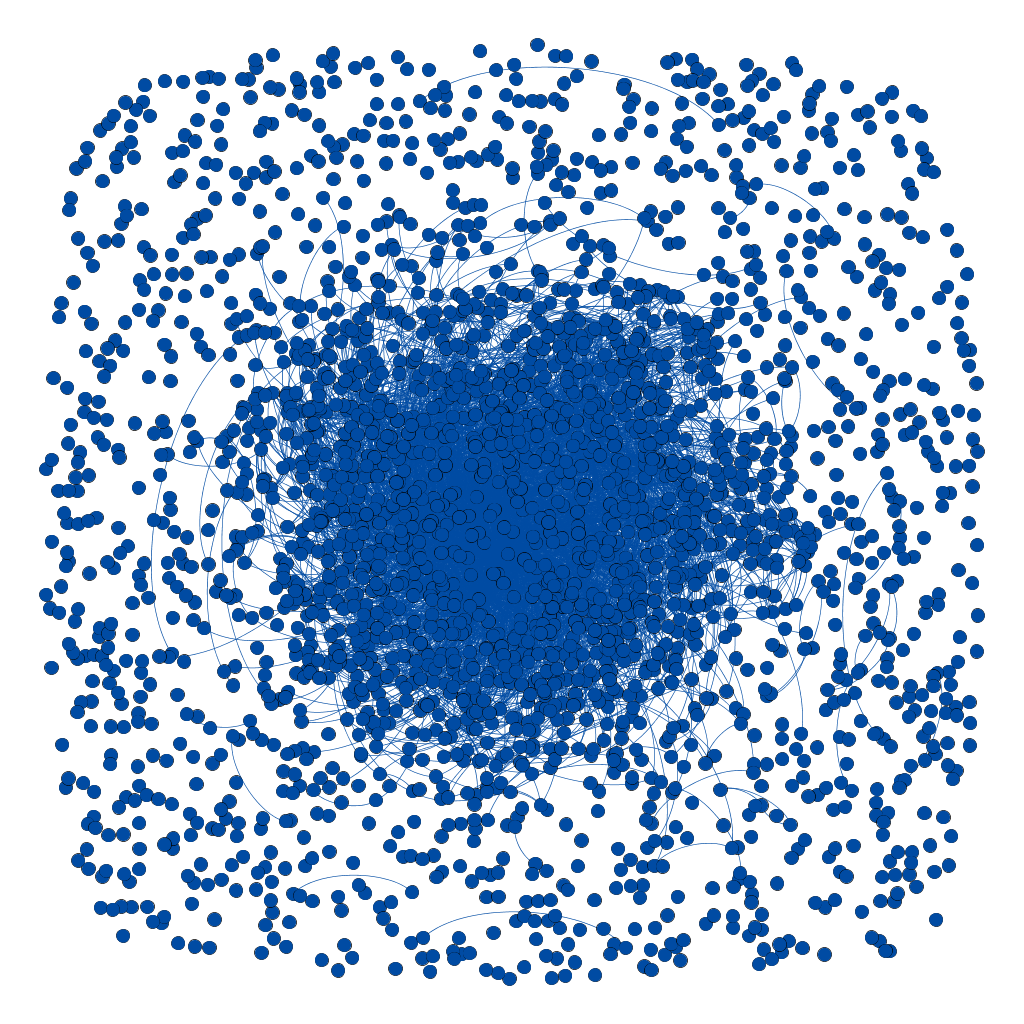}
        \caption{COVID-19 Test }
           \label{fig:core-peri-vax-test}
    \end{minipage}%
    \begin{minipage}{.25\textwidth}
      \includegraphics[width=5cm, height=3.5cm]{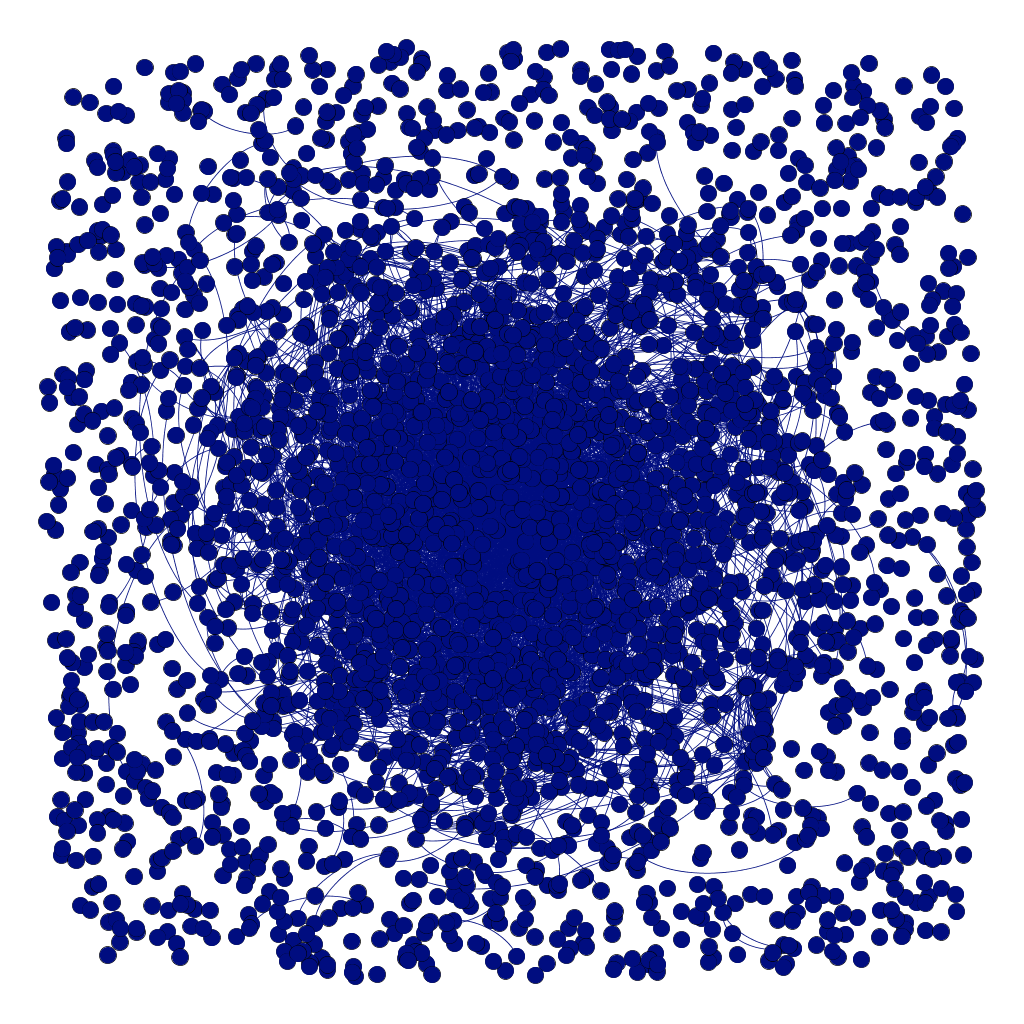}
      \centering
        \caption{Economic issues}
           \label{fig:core-peri-vax-eco}
    \end{minipage}%
 \begin{minipage}{.25\textwidth}
      \includegraphics[width=5cm, height=3.5cm]{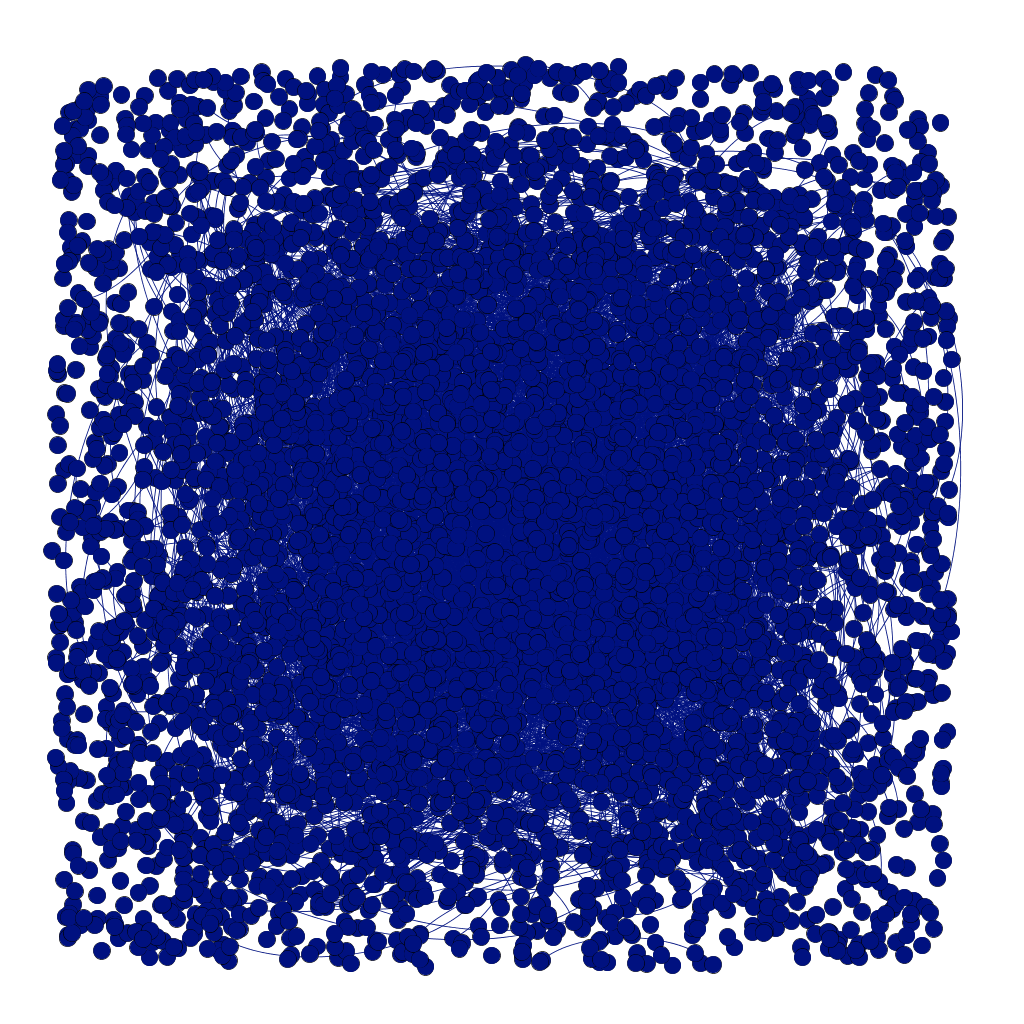}
        \caption{Random Graph}
           \label{fig:core-peri-vax-random}
    \end{minipage}%
     \centering
     \caption{Core and Periphery visualization of the Vaccine Data set}
    \label{fig:core-peri-vax}
\end{figure*}
\subsection{Result Analysis}
\label{sec:analysis}
\textbf{Kamala Harris Network}: Figure \ref{fig:result-pipline-harris} represents Kamala Harris network. This network is interesting because even in the context of the Capitol Attack and the Senate approval of election results, it is dominated by \#ADOS conversations, not by the other political issues including economic policies. An analysis of the three interesting measures for this network reveals the following.
The Eigenvector Centrality Disparity for the hashtag network (Figure \ref{fig:result-pipline-harris-ht-egc}) shows that all the groups generated by the subqueries are equally distributed, while the random graph has a higher volume with similar trends. Hence, these three subqueries are equally important. However, there are a few spikes on \#ADOS and “Capitol Attack” that indicates the possibility of interestingness. Figure \ref{fig:result-pipline-harris-ht-nav} shows that most of the “economic issues” has a spike on lower centrality nodes, but it touches zero with the increased centrality. While “Capitol attack” and “\#ADOS” has many more spikes in different centrality level. Hence, we conclude that this network is much more navigable for  “Capitol attack” and “\#ADOS”. However, the figure \ref{fig:result-pipline-harris-ht-prop} represents Propagativeness Disparity, and it is clear from this picture that African American issues dominated the conversation here, any subtopic related to it including non-US issues like ``Tigray'' (on Ethiopian genocide) propagated quickly in this network.  
\\
\noindent
\textbf{Biden Network}: While the predominance of \#ADOS issues might still be expected for the Kamala Harris data set, we discovered it to be an ``dominant enough'' subgraph in the Joe Biden data set represented in figure \ref{fig:result-pipline-biden}. The eigenvector centrality disparity shows that the three subgroups are equally dominated in this network. The “Joe Biden” data set represented in figure \ref{fig:result-pipline-biden-ht-egc}. The eigenvector centrality disparity shows that the three subgroups are equally dominated in this network. However, the navigability of the network(figure \ref{fig:result-pipline-biden-ht-nav}) also shows that it is navigable for all three subgraphs. It has two big spikes for “economic issues” and “Capitol attack,” plus many mid-sized spikes for "\#ADOS". Interestingly, in the figure \ref{fig:result-pipline-biden-ht-prop} the propagativeness shows that the network is strongly propagative with the economic issues and the \#ADOS issue, which shows up both in the Capitol Attack and the ADOS subgroups. Interestingly, the “Joe Biden” data’s co-mention network shows more propagativeness than the hashtag co-occur network, which indicates exploring the co-mention subgraph will be useful. We also note the occurrence of certain complete unexpected topics (e.g., COVIDIOT, AUSvIND -- Australia vs. India) within the ADOS group, while Economic Issues for Biden do not exhibit surprising results.
\\
\noindent
\textbf{Vaccine Network} In the vacation network, we found that “economic issues” and “covid tests” are more propagative than  “vaccine and anti-vaccine” related topics (Figure \ref{fig:result-pipline-vax}). The surprising result here is that the “vaccine - anti-vaccine” topics show a strong correlation with “Economy” in the other two charts. We observe that while the vaccine issues are navigable through the network, but this topic cluster is is not very propagative in the network.  In contrast, in the co-mention network, vaccine and anti-vaccine issues are both very navigable and strongly propagative. Further, the propagativeness in the co-mention network for the Covid-test shows many spikes at the different levels, which signifies for testing related issues, the network serves as a vehicle of message propagation and influencing. 

\subsection{Result Validation}
\label{sec:validation}

There is not a lot of work in the literature on interesting subgraph finding. Additionally, there are no benchmark data sets on which our ``interestingness finding'' technique can be compared to. This prompted us to evaluate the results using a core-periphery analysis as indicated earlier in Figure \ref{fig:sparse}. The idea is to demonstrate that the parts of the network claimed to be interesting stand out in comparison to the network of a random sample of comparable size from the background graph. These results are presented in Figures \ref{fig:core-peri-kamal}, \ref{fig:core-peri-biden}, \ref{fig:core-peri-vax}. In each of these cases, we have shown a representative random graph in the rightmost subfigure to represent the background graph. To us, the large and dense core formation in Figure \ref{fig:core-peri-kamal}(b) is an expected, non-surprising result. However, the lack of core formation on Figure \ref{fig:core-peri-kamal}(c) is interesting because it shows while there was a sizeable network for economics related term for Kamala Harris, the topics never ``gelled'' into a core-forming conversation and showed little propagation. Figure \ref{fig:core-peri-kamal}(a) is somewhat more interesting because the density of the periphery is far less strong than the random graph while the core has about the same density as the random graph. The core periphery separation is much more prominent in first three plots of Figure \ref{fig:core-peri-biden}. Unlike the Kamala Harris random graph, Figure \ref{fig:core-peri-biden}(d) shows that the random graph of this data set itself has a very large dense core and a moderately dense periphery. Among the three subfigures, the small (but lighter) core of Figure \ref{fig:core-peri-biden}(b) has the maximum difference compared to the random graph, although we find \ref{fig:core-peri-biden}(a) to be conceptually more interesting. For the vaccine data set, Figure \ref{fig:core-peri-vax} (c)is closest to the random graph showing that most conversations around the topic touches economic issues, while the discussion on the vaccine itself (Figure \ref{fig:core-peri-biden}(a)) and that of COVID-testing (Figure \ref{fig:core-peri-biden}(b)) are more focused and propagative. 

\section{Conclusion}
\label{sec:conclusion}
In this paper, we presented a general technique to discover interesting subgraphs in social networks with the intent that social network researchers from different domains would find it as a viable research tool. The results obtained from the tool will help researchers to probe deeper into analyzing the underlying phenomena that leads to the surprising result discovered by the tool. While we used Twitter as our example data source, similar analysis can be performed on other social media, where the content features would be different. Further, in this paper, we have used a few centrality measures to compute divergence-based features -- but the system is designed to plug in other measures as well.  
\FloatBarrier
\begin{acknowledgements}
This work was partially supported by NSF Grants \#1909875, \#1738411. We also acknowledge SDSC cloud support, particularly Christine Kirkpatrick \& Kevin Coakley, for generous help and support for collecting and managing our tweet collection.
\end{acknowledgements}
\bibliographystyle{spbasic}     
\bibliography{refs}   

\end{document}